\UseRawInputEncoding
\documentclass[aps, preprint, letterpaper, amsmath, amssymb, nofootinbib, longbibliography]{revtex4-2}
\makeatletter
\def\active@comma{,}
\makeatother
\usepackage{xcolor}
\usepackage{tabularray}
\usepackage{graphicx}% Include figure files
\usepackage{dcolumn}% Align table columns on decimal point
\usepackage{bm}% bold math
\usepackage[a4paper, total={6in, 9in}]{geometry}
\usepackage{graphicx}
\usepackage{amssymb}
\usepackage{amsthm}
\usepackage{bbm}
\usepackage{bm}
\usepackage{soul}
\usepackage{amsmath}
\usepackage{mathrsfs}
\usepackage{amssymb}
\usepackage{exscale}
\usepackage{easybmat}
\usepackage[T1]{fontenc}
\usepackage[utf8]{inputenc}
\usepackage{xcolor}
 
\usepackage{comment}
\usepackage[normalem]{ulem}
\usepackage{booktabs}
\usepackage{multirow}
\usepackage{siunitx}
\usepackage[figurename=Figure, labelsep=period, labelfont=bf, format=plain]{caption}
\usepackage{titlesec}
\usepackage{mathtools}
\usepackage{float}
\usepackage{tabularx}
\usepackage{adjustbox}
\usepackage{lipsum}
\usepackage{bbm}
\usepackage[dvipsnames]{xcolor,colortbl}
\usepackage{amsmath,mathrsfs,lineno,amssymb,amsbsy}
\usepackage{hyperref}
\usepackage{upgreek}
\usepackage{bm}
\usepackage[normalem]{ulem} % [normalem] prevents the package from changing \emph to underline
\UseTblrLibrary{booktabs}
\SetTblrStyle{note}{
    halign=l
}
\usepackage{tikz}

\definecolor{wn}{HTML}{7B1E78}       % wall-normal
\definecolor{sw}{HTML}{168C9C}       % streamwise
\definecolor{stabcolor}{HTML}{1746D1}     % stabilization
\definecolor{destabcolor}{HTML}{B02018}   % destabilization

\definecolor{ruleone}{HTML}{F0CF3D}
\definecolor{ruletwo}{HTML}{D7A20C}
\definecolor{rulethree}{HTML}{B87300}
\definecolor{rulefour}{HTML}{8C4A00}

\newcommand{\RuleBadge}[2]{%
  \tikz[baseline=(n.base)]\node[
    circle,
    fill=#1,
    draw=black,
    line width=0.5pt,
    text=white,
    inner sep=1.7pt,
    minimum size=5.5mm,
    font=\bfseries\small
  ] (n) {#2};%
}

\usepackage{array}

\renewcommand\thesection{\arabic{section}}
\renewcommand\thesubsection{\thesection.\arabic{subsection}}

\makeatletter
\def\p@subsection{}
\def\p@subsubsection{}
\makeatother

\newlength{\savedtextwidth}
\begin{document}

\title{A deep dive into Tollmien-Schlichting wave control via passive wall deformations: The battle between local and downstream stabilization, lessons learned, and implications for \\ phononic subsurfaces}
\date{\today}

\author{H. Yousef$^{1,\dagger}$}
\author{H. Hassan$^{1,\dagger}$}
\author{I. Roy$^{2}$}
\author{T. Toki$^{2,3}$}
\author{C. Scalo$^{2,3}$}
\author{M. Nouh$^{1,4}$}\altaffiliation[Corresponding author]{(mnouh@buffalo.edu)}

\affiliation{%
\vspace{2ex}%
\centering
\footnotesize
\setlength{\baselineskip}{0.80\baselineskip}
{\normalfont $^1$}Dept. of Mechanical and Aerospace Engineering, University at Buffalo (SUNY), Buffalo, NY 14260-4400, USA\\[1ex]
{\normalfont $^2$}School of Mechanical Engineering, Purdue University, West Lafayette, IN 47907-2088, USA\\[1ex]
{\normalfont $^3$}School of Aeronautics and Astronautics, Purdue University, West Lafayette, IN 47907-2045, USA\\[1ex]
{\normalfont $^4$}Dept. of Civil, Structural and Environmental Engineering, University at Buffalo (SUNY), Buffalo, NY 14260-4300, USA\\[1ex]
{\normalfont $^\dagger$}These authors contributed equally to this work
}

\vspace{1cm}

\begin{abstract}
\noindent \rule{\linewidth}{0.5pt} 
\normalsize

A decade ago, a landmark study on flow control via subsurface phonons transformed our understanding of fluid-structural interactions, compelling us to reimagine ways by which to suppress boundary layer instabilities. While subsequent investigations have steadily enriched this landscape, several questions remain largely unanswered. The notion of Tollmien-Schlicting (TS) wave stabilization relies on phase-engineered surface interactions, which destructively engage with the wave and impede its growth. Although the fundamental drivers of the phenomenon are established, its granular physics remain insufficiently resolved, particularly as it pertains to directional effects, competing energy production mechanisms, and the precise streamwise locations governing the process. Revisiting this initial vision, we confront these open questions, revealing fresh insights, and establishing broad foundational strokes to guide the next era of investigations. We begin by defining the amplitude and phase criteria of an elastic wall admittance driving perturbation energy changes relative to a rigid wall. The established framework isolates the roles of the work-rate, viscous energy, production, and dissipation in shaping the fluid’s response and clarifies how each contributes to the collective outcome. Crucially, we demonstrate how streamwise translation of the interaction surface significantly alters these parameters, revealing critical thresholds at which the control result is fully reversed. Our model identifies specific pathways to sustained TS wave attenuation, and defines the limits of what can be accomplished passively. Finally, we conceptualize a two-dimensional phononic subsurface, whose tailored and disproportional response to the flow in the wall-normal and streamwise directions, brings about the elusive combination of \mbox{local and downstream stabilization}.  

\noindent \rule{\linewidth}{0.5pt} 

\end{abstract}
\maketitle

%%%

%%%%%%%%%%%%%%%%%%%%%%%%%%%%%%%%%%%%%%
\newpage
\section{Introduction \label{sec:Intro}}
%%%%%%%%%%%%%%%%%%%%%%%%%%%%%%%%%%%%%%

Flow control is ubiquitous in nature. Flying and swimming organisms have long been to known to manipulate the surrounding fluid to modify its flow characteristics, consequently generating lift, propulsion, and maneuverability \cite{Fish2008, Guo2025, Fish2026}. These naturally occurring flow control strategies have inspired analogous approaches in engineered systems, in which boundary layer dynamics play a central role. The pioneering work of Prandtl established the theoretical framework for boundary layer flows and laid the foundation for modern flow control research \cite{LudwigPrandtl1904}. Since then, extensive research has investigated flow control techniques across a broad range of applications, with particular emphasis on boundary layer separation \cite{Lin2002}, flow-induced noise \cite{Geyer2009}, and laminar flow instabilities \cite{Liepmann1982, Kachanov1994}. Many of these efforts have opened up new avenues to reduce aerodynamic drag, enhance lift, boost thrust generation, and enable \mbox{quieter flight}.

In this work, we focus on the long-standing problem of laminar-to-turbulent transition of a subsonic boundary layer developing over a flat plate. Under low disturbance conditions, this process is typically instigated by primary viscous instabilities, commonly known as Tollmien-Schlichting (TS) waves, manifesting as coherent fluctuations in the flow field that amplify within the boundary layer, ultimately triggering secondary instabilities and accelerating the breakdown to turbulence. The increased skin friction drag, and the decreased aircraft efficiency in the aftermath of these waves, have well-established detrimental engineering and economic implications on aerodynamic systems which have been studied for decades \cite{Hefner1988, ren2020aviation}. Inspired by the wave-like nature of TS instabilities suggested by the Orr-Sommerfeld equation, early experimental work by Milling demonstrated successful wave cancellation in a water channel using a vibrating wire tuned out-of-phase to induce destructive interference \cite{Milling1981}. Several numerical and experimental investigations have subsequently followed suit, exploring active techniques such as plasma actuators \cite{Grundmann2008, Kotsonis2013, Kotsonis2015}, distributed blowing and suction \cite{Danabasoglu1991}, and surface heating and cooling \cite{Brennan2020}, to mitigate TS waves. Despite showing promise, these active methods face significant constraints, notably the high actuation energy, which often negates the net drag reduction gains. Furthermore, practical deployment remains greatly hindered by the requirement of bulky sensors to detect flow properties and the need for complex processing algorithms for real-time actuation \cite{Walther2001, Mohammadikalakoo2024}.

In the face of these limitations, passive solutions remain more appealing and practically potent. Yet, the limited versatility of a passive flow control mechanism, by definition, requires radical and deeply innovative concepts in pursuit of the same outcomes achieved through active means. For instance, passive porous coatings, modeled as thin perforated plates, were initially shown to completely stabilize TS waves via targeted phase values \cite{Carpenter2001}. However, subsequent studies incorporating Darcy's law to model flow within the porous matrix revealed an earlier onset of transition in channel flows \cite{Tilton2006}. Experimental sea trials by Kramer, using deformable compliant coatings inspired by dolphin skin, reported significant drag reduction \cite{Kramer1960}. And while initial follow-on experiments failed to reproduce the same level of compliant wall effectiveness \cite{Bushnell1977}, the pivotal work of Gaster revived the interest in compliant walls a decade later \cite{Gaster1988}, prompting extensive theoretical and computational investigations into the effect of compliant walls on growth rate of flow instabilities \cite{Lucey1995, Gad-El-Hak1986, Malik2018}. For example, a compliant finite wall was shown to stabilize TS waves \cite{Yeo1988, Davies1997}, providing an attractive avenue for passive transition delay. Nevertheless, these configurations remain susceptible to surface fluid-structure instabilities induced by large wall deformations, commonly referred to as traveling-wave flutter (TWF). Furthermore, high structural compliance is generally incompatible with load-bearing structures such as aircraft wings, where boundary layer transition typically develops \cite{Carpenter2001-2}.

Although fluid-structure interactions is a mature discipline, research efforts have historically focused on fluid flows interacting with conventional structures or geometric wall modifications (e.g., riblets and grooves \cite{choi2013smart}). Over the past three decades, however, rapid advancements in architected material physics have fundamentally changed our understanding of wave-matter synergy, unlocking new paradigms in wave scattering, dispersion, directionality, and frequency-dependence \cite{attarzadeh2020experimental, moghaddaszadeh2021nonreciprocal, hu20203d, moghaddaszadeh2023local, mousa2024parallel}. The ability of such ``metamaterials'' to exhibit an engineered elastodynamic response stems from their internal architecture and the spatial arrangements of their internal building blocks, which trigger bulk properties and wave propagation characteristics, that go beyond what their constitutive units can solely accomplish \cite{Chaplain2025, Duncan2026}. Recently, with the interplay between fluid dynamics and metamaterials as a driving theme, a new research community has started to take shape \cite{Krushynska2025EUROMECH659, Matlack2026FMI}. Under this umbrella, recent studies have increasingly explored the potential of harnessing metamaterials for advanced fluid manipulation. Notable advancements include noise mitigation in acoustic cavities \cite{aladwani2019fluid}, the characterization of nonlinear coupled interactions in aerodynamic flows \cite{Ramakrishnan2026}, analyzing two-way interactions between wall-bounded turbulence and defect-embedded structures \cite{Lin2026}, and extracting undesirable fluid perturbations via acoustic diodes \cite{Schmidt2025}. More recent applications have broadened the reach of this research direction, ranging from separated flows \cite{Ramakrishnan2026-2} and TWF in compliant materials \cite{Fabbiane2025}, to shockwave mitigation \cite{Navarro2025}, as well as metasurface-induced \cite{Zhao2022} and phonon-mediated \cite{Brehm2026} hypersonic boundary layer stabilization. A recent comprehensive review of the majority of research efforts in this space can be found in \cite{Avallone2026}.

With the central scientific premise of harmonizing the spatial and temporal scales of these architected materials with those of a given fluid flow, Hussein et al. demonstrated the ability of a passive phononic crystal placed beneath the surface of a water channel to achieve localized TS wave attenuation, as a direct result of its tuned phase- and frequency-dependent properties \cite{Hussein2015}. The seminal work of this team introduced to the community the notion of a phononic subsurface (PSub), a material whose tailored vibrational response serves as a blueprint for flow control. A second investigation demonstrated the indifference of the core idea to the type of the PSub, achieving equally-powerful results with locally resonant metamaterials in lieu of phononic crystals \cite{Kianfar2023-2}. While intriguing, both studies identified a number of critical shortcomings in the underlying approach, most importantly, the narrowband nature of the control mechanism (occurring largely in the small vicinity of a resonant frequency) and the confinement of the TS wave attenuation to the fluid-PSub interaction region. Specifically, downstream of this region, a substantial rebounding effect causes the TS wave to behave demonstrably worse with the PSub than without it. Subsequent studies consistently confirmed these limitations \cite{Barnes2021}, and further revealed additional insights into the physics underpinning the two-way interactions between the unstable flow and the PSub; most notably, competing perturbation kinetic energy behaviors for the wall-normal and the streamwise velocity components \cite{Michelis2023}. Early experimental efforts also provided valuable insight \cite{Keogh2025}, despite the extremely challenging nature of creating the conditions necessary for PSub evaluation in a wind tunnel environment \cite{juhl2026lessons}. In pursuit of solutions, complementary mechanisms have since been proposed that extend the interaction domain along the fluid surface \cite{Kianfar2023} and broaden the bandwidth of the resonant modes \cite{Harris2026}. On a separate path, other efforts have tackled the downstream recovery problem by employing multi-interaction systems, in which the flow interfaces with the same PSub at two locations, with the goal of achieving phasing conditions that are otherwise unattainable via a single passive structure \cite{Willey2023}, or by implementing a lattice of PSubs that influence the amplified downstream perturbations through scatterless interferences \cite{Hussein2026}. 

Despite the surging interest in the problem, the ability to achieve sustainable attenuation of TS waves via phase-synchronized wall motion remains an open question. While valiant, the aforementioned attempts provide a correction mechanism rather than a root cause solution to the single, passive PSub idea. The partially understood nature of the problem, combined with this elusive combination of localized and downstream stabilization, motivate the need for this study. In here, we set out to achieve two goals: The first is to dissect the coupled flow-PSub problem by diving deeply into the amplitude and phase conditions that drive the perturbation energy changes, relative to a rigid wall. We establish a robust predictive framework based on a generalized mechanical admittance boundary condition, and verified by Linear Stability Theory (LST). The model breaks down the different contributors to these changes, from the interface work-rate and viscous energy terms to production and viscous dissipation. We explore the roles played by each of these mechanisms, and investigate the different ways in which they shape the fluid's response, along with their sensitivity to the driving velocity component at the boundary and the interaction location. In doing so, we reduce reliance on assumptions, by establishing an evidence-based coupled framework which is no longer flow agnostic. Secondly, we build on this nuanced understanding of the problem to establish a novel pathway towards achieving concurrent localized and downstream stabilization. We do so by introducing the concept of a two-dimensional PSub, which invokes the appropriate control knobs influencing TS wave growth at both locations, guided by our evolved interpretation of the different underlying interaction mechanisms. The framework shown here brings the core PSub concept one step closer to practical implementation. More broadly, research bridging fluid dynamics and architected materials could revolutionize flow control, unlocking new applications in biomedical devices, wind energy, microfluidics, and smart aerodynamics.

%%%%%%%%%%%%%%%%%%%%%%%%%%%%%%%%%%%%%%
\section{Computational setup} \label{sec:Comp set}

The problem under investigation is of the development and control of Tollmien-Schlicting (TS) waves in a zero-pressure boundary layer flow over a flat plate. Figure~\ref{fig:Comp Setup}a illustrates the fluid computational domain. We adopt a similar setup to that used in Refs.~\cite{Michelis2023, Michelis20232}, where a subsonic free stream velocity $U_\infty =24$ [m/s] is imposed at the inlet with air as the operating fluid at Standard Temperature and Pressure (STP) having a density $\rho_f = 1.20$ [kg/m$^3$], dynamic viscosity $\mu_f = 18.13$ [$\mu$Pa$\cdot$s], freestream pressure $P_\infty =1$ [MPa] and freestream temperature $T_\infty = 293$ [K]. The computational domain extends to a height $h_f =0.3$ [m] in the wall-normal $y$-direction and spans a length $\ell_f=1.6$ [m] along the streamwise $x$-direction. The flat plate itself (bottom wall of the computational domain) is $\ell_p = 1$ [m] long with a zero velocity no-slip boundary condition leading to the formation of the boundary layer due to viscous effects. An inlet domain $\ell_i = 0.3$ [m] long with the freestream velocity imposed on the left side of the domain resolves the leading edge of the flat plate, while an outlet domain $\ell_o =0.3$ [m] long covers the right side of the flat plate with an outlet boundary condition at the rightmost edge. The bottom boundaries of the inlet and outlet domains have a symmetry condition, whereas the top of the computational domain has an open boundary condition mimicking the freestream flow. 

\subsection{Numerical framework} \label{sec:Num Frame}

\begin{figure*}[h!]
    \includegraphics[width=1\textwidth]{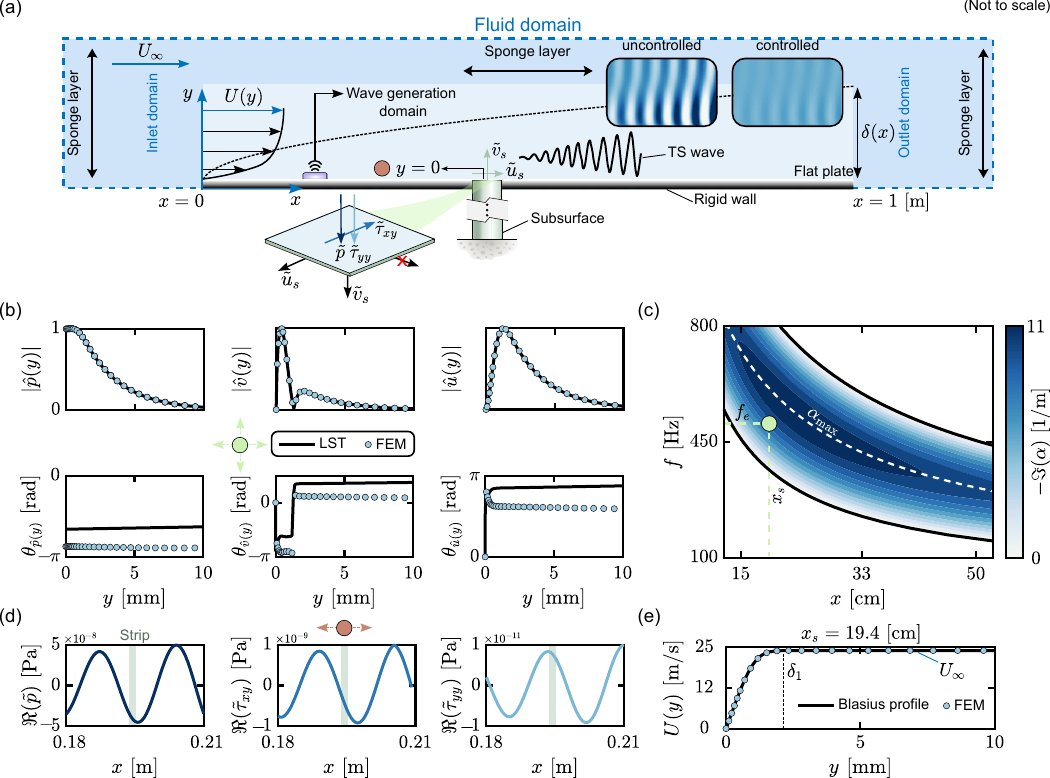}
    \caption{\textbf{Computational setup and baseline assessment of TS wave properties for the reference (rigid-wall) case.} (a) Schematic of the computational domain. A $1$ [m] flat plate is placed beneath the flow starting at $x=0$ [m], with a free stream velocity, $U_\infty$, and inlet and outlet domains on both sides. Wave generation (forcing) domain and sponge layers are labeled accordingly. A hypothetical subsurface structure is also defined for future discussions. (b) Mode shape amplitude and phase of the perturbation field parameters, $\hat{p}$, $\hat{u}$, and $\hat{v}$, obtained via FEM and confirmed by LST, at $f_e=500$ [Hz] and $x_s=19.4$ [cm]. (c) Neutral stability curves of the TS waves over the flat plate obtained from the LST analysis. The white dashed curve tracks the maximum growth and the circular marker indicates the coordinates of the $(f_e, x_s)$ point at which the mode shapes are extracted. (d) Variation of the perturbed pressure, $\tilde{p}$, and viscous stresses, $\tilde{\tau}_{xy}$ and $\tilde{\tau}_{yy}$, at the wall. The shaded region highlights the region of flow control via elastic wall motion, for future reference. (e) Boundary layer profile, $U(y)$, obtained via FEM (circular markers) and imposed on the Blasius solution (solid curve), with the boundary layer displacement thickness, $\delta_1$ (dashed vertical line), and freestream velocity, $U_\infty$, highlighted.}
    \label{fig:Comp Setup}
\end{figure*}

The fluid-structural coupling associated with this problem is modeled using the Finite Element Method (FEM). Simulations are performed in two steps: First, the two-dimensional incompressible Navier–Stokes (NS) equations are used to establish the steady boundary layer ($\partial \mathbf{u}/\partial t=0$) over the flat plate., i.e.,
\begin{subequations}
\begin{equation}
\frac{\partial \mathbf{u}}{\partial t}= - (\mathbf{u} \cdot \nabla )\mathbf{u}-\frac{1}{\rho_f}\nabla p + \frac{\mu_f}{\rho_f} \nabla^2 \mathbf{u} 
\label{eq:NS}
\end{equation}
\begin{equation}
\nabla \cdot \mathbf{u} =0
\label{eq:Continuity}
\end{equation}
\end{subequations}
where $\mathbf{u} =[u,v]$ and $p$ are the velocity and pressure fields, respectively, with $u$ and $v$ denoting the streamwise and wall-normal velocity components and the $\nabla$ is the vector differential operator. Secondly, the adiabatic Linearized Navier-Stokes (LNS) equations are solved for the velocity and pressure perturbation fields. Specifically, Eqs.~(\ref{eq:NS}) and (\ref{eq:Continuity}) are linearized around the base flow field, $\mathbf{U}=[U,V]$, yielding the LNS equations given by Eqs.~(\ref{eq:LNS_all}), 

\begin{subequations}
\begin{equation}
\frac{\partial \tilde{\mathbf{u}}}{\partial t} + (\mathbf{U}\cdot \nabla )\tilde{\mathbf{u}} + (\tilde{\mathbf{u}}\cdot \nabla ){\mathbf{U}}= -\frac{1}{\rho_f}\nabla \tilde{p} + \frac{\mu_f}{\rho_f} \nabla^2 \tilde{\mathbf{u}} 
\label{eq:LNS}
\end{equation}
\begin{equation}
\nabla \cdot \tilde{\mathbf{u}} =0
\label{eq:Continuity2}
\end{equation}
\label{eq:LNS_all}
\end{subequations}
with the perturbation field ($\tilde{\mathbf{u}},\tilde{p}$). The inlet and outlet domains along with a top segment of the computational domain are replaced with a sponge layer with boundaries matching the impedance of the fluid to perfectly transmit waves away from the computational domain as highlighted by the blue background in Fig.~\ref{fig:Comp Setup}a. The upper sponge layer starts at $y_{sl} =0.1$ [m] with a height of $h_{sl} = 0.2$ [m].

The simulations are carried out in $\textsc{Comsol Multiphysics}$ 6.3, implementing a combination of the Fluid Flow, Heat Transfer, and Acoustics modules. Linear elements are used as the basis for the FEM in step one, solving for the velocity and pressure fields. Simulation outcomes are then complemented with analytical results from a spatial Linear Stability Theory (LST) analysis which is discussed thoroughly in Sec.~\ref{sec:LST}. A volume force source $p_e$ taking the following form,
\begin{equation} 
p_e(x,y,t) = A_e\bigg[\sin^2\Big(\pi\Big[\frac{x-x_e}{\ell_e}\Big]\Big) \sin^2\Big(\pi\Big[\frac{y-y_e}{h_e}\Big]\Big)\bigg] ~ \cos(\omega_e t)
\label{eq:Excitation}
\end{equation}
is injected near the flat plate inside the boundary layer to excite the TS waves, where $x_e=0.1$ [m] and $y_e=0$ [m] denote the starting coordinates of the wave generation domain, $\ell_e=10$ [mm] and $h_e=0.5$ [mm] denote its length and height, respectively, while $A_e = 2 \times 10^{-5}$ [N/m$^3$] and $\omega_e=2\pi f_e$ are the forcing amplitude and forcing angular frequency, respectively. Quadratic Lagrange elements are used as the basis for the FEM in step two. Details of the meshing procedure and the convergence study are provided in Appendix~\ref{app:Mesh}. A monochromatic frequency-domain study with $f_e = 500$ [Hz] is carried out to establish the growth characteristics of the TS waves inside the boundary layer and compared against LST predictions. Figure~\ref{fig:Comp Setup}e shows excellent agreement between the boundary layer profile, $U(y)$, obtained from the FEM (circular markers) when compared with the Blasius solution (solid curve) at a streamwise location $x_s=19.4$ [cm] (A location which will later mark the starting point of the elastic portion of the wall when flow control is pursued). 

\subsection{Spatial Linear Stability Theory (LST) analysis} \label{sec:LST}

A spatial LST analysis is carried out to establish the frequency spectrum of the TS waves for a flat plate with a fully-rigid wall, and support the design of a positive intervention mechanism through fluid-structure interactions with a small section of the wall which is replaced with an elastic surface. The LST expresses the fluctuating flow variables as normal modes, $\tilde{\phi} =\hat{\phi}(y)e^{\mathrm{i}(\alpha x-\omega t)}$, which, when substituted into the linearized governing equations, yield an eigenvalue problem. Here, $\tilde{\phi} \in [\tilde{p}, \tilde{T}, \tilde{u}, \tilde{v}]$ represents any of the fluctuating variables, with $\alpha$ being the complex streamwise wavenumber and $\omega=2 \pi f$ being the real angular frequency. The LST analysis assumes the boundary layer to be locally parallel, restricting the mode shapes, $\hat{\phi}$, to depend only on the wall-normal coordinate, $y$. When the above ansatz is substituted into the linearized governing equations, the resulting system of equations takes the following form,
\begin{equation}\label{eqn:LST_eqn1}
    \mathbf{C}_1\frac{d^2 {\hat{\phi}}}{dy^2} + \mathbf{C}_2\frac{d {\hat{\phi}}}{dy} + \mathbf{C}_3 {\hat{\phi}} = 0
\end{equation}
with the matrices $\mathbf{C}_1$, $\mathbf{C}_2$, and $\mathbf{C}_3$ representing the coefficient matrices. In these matrices, each column is associated with the corresponding fluctuation terms and each row represents the coefficients in each of the governing equations, i.e., the conservation equations for mass, momentum, and energy. Our in-house developed LST solver is developed for the full compressible Navier-Stokes equations for applicability to different Mach number regimes. The solver has been predominantly used for supersonic and hypersonic base flows in our research group \cite{Roy2025}, and has been validated against experimental studies \cite{Miller2026, Sousa2023}. Implementation details on the Laguerre spectral method and relevant boundary conditions are given in Ref.~\cite{Roy2026}. 

\newpage
In this work, we deploy this LST solver to a low Mach number base flow. The TS wave perturbation fields, $\tilde{p}$, $\tilde{u}$, and $\tilde{v}$, at $x_s=19.4$ [cm] are obtained from both the FEM and the LST for the reference (rigid-wall) case, and are presented in Fig.~\ref{fig:Comp Setup}b. The top panel of the figure compares the amplitudes of each field quantity (normalized by the maximum value) with excellent matching between the two approaches. The bottom panel show the phase of each component. While FEM is able to capture the phase trend for the entire field, the deviations from LST predictions is expected and can be attributed to the presence of non-parallel effects in the FEM simulations as the wall-normal base velocity ($V\neq0$) in contrast to the locally parallel LST calculations, which enforce a zero base wall-normal velocity. It is worth noting that the phase difference shown here plays an important role throughout the discussions, since the fundamental TS mode amplification mechanism relies on the phase relationship between $\tilde{u}$, $\tilde{v}$, and $\tilde{p}$.

Figure~\ref{fig:Comp Setup}c shows the neutral stability curve of the TS waves over the flat plate, obtained from the LST analysis, with $-\Im(\alpha)$ denoting the growth rate. TS mode amplification takes place over the $x \approx 14$ [cm] to the $x \approx 44$ [cm] range. The dashed curve represents the maximum amplification line within the growth zone, $\alpha_{\mathrm{max}}$, which roughly captures the evolution of the primary TS mode as it propagates downstream. In the figure, $f_e$ is indicated by the horizontal dashed line, while $x_s$ is indicated by the vertical dashed line. The combination of the two is highlighted by the green circular marker, which corresponds to a moderately-amplified TS mode.

\subsection{Fluid-structure interaction} \label{sec:FSI}

After establishing the properties of the TS waves developing in the subsonic boundary layer and casting the LNS equations, Eqs.~(\ref{eq:LNS}) and (\ref{eq:Continuity2}), in the frequency domain, a localized section of the rigid wall is replaced with an elastic surface (starting at $x_s$), enabling fluid-structure interaction that can be tuned to modulate and potentially suppress the TS wave growth. In practice, this surface can be realized via a small elastic wall strip, or a full structure extruding below the surface of the flow (commonly known as a subsurface structure). We will begin with the former scenario in order to extract critical insights into the underlying interaction mechanisms, and then move on to the latter to discuss broader practical considerations. As such, consider an $\ell_s =1$ [mm] wide segment of the rigid wall to be replaced with an elastic strip, starting at $x_s=19.4$ [cm]. We start by modeling this strip via a generalized boundary condition given by the admittance matrix,
\begin{equation}
\mathbf{Y}
=
\begin{bmatrix}
{Y}_{xx}(\omega) & {Y}_{xy}(\omega)\\
{Y}_{yx}(\omega) & {Y}_{yy}(\omega)
\end{bmatrix}
\label{eq:admittance matrix}
\end{equation}

The fluid Cauchy stress tensor is $\boldsymbol{\sigma}_f = -\tilde{p} \mathbf{I} + \tilde{\boldsymbol{\tau}}$, where $\mathbf{I}$ is the identity matrix and $\tilde{\boldsymbol{\tau}}$ is the perturbed viscous stress tensor for a Newtonian fluid, given by
\begin{equation}
\tilde{\boldsymbol{\tau}} = \mu_f \Big[\nabla \tilde{\mathbf{u}} + \left(\nabla \tilde{\mathbf{u}} ^T\right)\Big]
\label{eq:viscous stress}
\end{equation}
with $\mathbf{n}_f$ denoting the outward unit vector normal to the fluid, pointing downward toward the wall, the traction exerted by the fluid on the structure follows as
\begin{equation}
\tilde{\mathbf{t}}_f
=-\boldsymbol{\sigma}_f\cdot\mathbf{n}_f
=
\begin{bmatrix}
\tilde{\tau}_{xy}(\omega)\\
-\tilde{p}(\omega)+\tilde{\tau}_{yy}(\omega)
\end{bmatrix}
\label{eq:tf_traction}
\end{equation}
where $\tilde{t}_{f_{xx}}=\tilde{\tau}_{xy}$ and $\tilde{t}_{f_{yy}}=-\tilde{p}+\tilde{\tau}_{yy}$ denote the streamwise
and wall-normal components of the fluid traction, respectively \cite{Gabard2020, Jafari2023}. To facilitate interpretation from the flow side, we define 
\begin{equation}
\tilde{\mathbf{q}}_f
=
\begin{bmatrix}
\tilde{\tau}_{xy}(\omega)\\
\tilde{p}(\omega)-\tilde{\tau}_{yy}(\omega)
\end{bmatrix}
\label{eq:fluid_load}
\end{equation}
to be the generalized load vector at the interface. Setting the perturbed velocity vector along the surface of the admittance strip, $\tilde{\mathbf{u}}_s =[\tilde{u}_s \,\,\, \tilde{v}_s]^T$, to be linearly dependent on $\tilde{\mathbf{q}}_f$ via the admittance matrix, yields the admittance boundary condition at the strip given by
\begin{equation}
\tilde{\mathbf{u}}_s = \mathbf{Y} ~ \tilde{\mathbf{q}}_f
\label{eq:admittance BC}
\end{equation}
Since all quantities in Eq.~(\ref{eq:admittance BC}) are frequency dependent, $(\omega)$ is henceforth omitted for brevity. Figure~\ref{fig:Comp Setup}d shows the perturbed pressure and viscous stresses along the streamwise direction at the fluid-structural interface ($y=0$) for the reference case, revealing comparable amplitudes of the pressure and shear viscous stress with small variations along the strip location. Variations in the normal viscous stress, however, are significantly smaller, i.e., $\tilde{\tau}_{yy} \ll \tilde{p}$, and it is therefore neglected. Consequently, the wall-normal admittance can be defined as $Y_{yy}=\tilde{v}_s/\tilde{p}$ from the flow side.

While Fig.~\ref{fig:Comp Setup}a depicts a generalized two-dimensional subsurface structure with the close-up inset showing the corresponding traction and velocity components of the top surface, any structural design can be encapsulated within the generalized admittance boundary condition defined by Eq.~(\ref{eq:admittance matrix}), with $Y_{yy}=\tilde{v}_s/\tilde{p}$ being the wall-normal admittance relating the pressure to the wall-normal velocity \cite{Tian2024}. For example, the upper diagonal term, $Y_{yy}$, is used to model passive porous walls. It is also used to depict blowing and suction in the absence of $Y_{yx}$, which yields a transpiration boundary condition, i.e., $\tilde{v}_s \neq 0$ with $\tilde{u}_s =0$ \cite{Tilton2015}. Modifying the previous condition by accounting for small deformations in the wall-normal direction, using a Taylor series linearization, results in a coupling between the velocity components of the boundary condition ($Y_{yx} \neq 0 \to \tilde{u}_s \neq 0 $), which is used to model compliant walls \cite{Davies1997} as well as phononic subsurfaces \cite{Hussein2015}. The cross terms, $Y_{xy}$ and $Y_{yx}$, are referred to as the momentum transfer admittances which relate the streamwise traction to wall-normal velocity, and the wall-normal traction to streamwise velocity, respectively \cite{Gabard2020}. These can be realistically implemented via angled perforated plates and anisotropic compliant walls \cite{Carpenter1990}. Finally, we note that the streamwise admittance term, $Y_{xx}$, is a friction-like term mimicking surface roughness, and is used to relate the streamwise traction to the streamwise velocity \cite{Aurgan2018}. 

While the admittance matrix in Eq.~(\ref{eq:admittance BC}) is not necessarily symmetric nor Hermitian, it is important to address the passivity of the admittance strip, i.e., the ability of the strip to receive more energy than it provides to the fluid through the work done by traction forces. The passivity of the wall-normal term, $Y_{yy}$, is explored in the absence of the remaining terms, as most commonly studied in literature \cite{Luhar2015, Song2026}. A two-way coupling is enforced at the interface between the fluid and the admittance strip, assuming one-dimensional vibrations along the wall-normal direction, resulting in Eqs.~(\ref{eq:traction cont}) and (\ref{eq:velocity cont}),
\begin{subequations}
\begin{equation}
\tilde{t}_{f_{yy}} = -\tilde{t}_{s_{yy}} 
\label{eq:traction cont}
\end{equation}
\begin{equation}
{\tilde{v}}_f = \tilde{v}_s
\label{eq:velocity cont}
\end{equation}
\label{eq:Two way coupling}
\end{subequations}
where $\tilde{v}_f$ is the perturbed fluid velocity at the interface, and $\tilde{t}_{s_{yy}}$ is the structural traction acting on the fluid, thereby simplifying Eq.~(\ref{eq:admittance BC}) to $\tilde{v}_s = Y_{yy} ~ \tilde{p}$. A single degree of freedom (SDoF) mass-spring-damper system is used to illustrate the coupling between the domains,
 \begin{equation}
\left(-\omega^2m +\mathrm{i}\omega c + k \right) \eta=\tilde{t}_{f_{yy}}
\label{eq:MS}
\end{equation}
where $m$, $c$, and $k$ are the mass, damping and stiffness coefficients, respectively, $\eta$ is the wall-normal displacement in the frequency domain. The structural admittance can be extracted from Eq.~(\ref{eq:MS}), yielding $Y_{yy} = -i\omega \eta/ \left(-\omega^2m +\mathrm{i}\omega c + k \right) \tilde{p}$, with the admittance phase falling within the range $\theta_{Y_{yy}} \in [\pi/2,3 \pi/2]$. The passivity constraint of the SDoF can be attributed to the system's rate of work at the interface, given by
\begin{subequations}
\begin{equation}
\mathcal{W}_s = \tilde{v}_s  ~ \tilde{t}_{f_{yy}}
\label{eq:work s}
\end{equation}
\begin{equation}
\mathcal{W}_f = \tilde{v}_f ~ \tilde{t}_{s_{yy}}
\label{eq:work f}
\end{equation}
\end{subequations}
Substituting the traction coupling condition given by Eq.~(\ref{eq:traction cont}) in the fluid work-rate relation given by Eq.~(\ref{eq:work f}), we conclude that $\mathcal{W}_s = -\mathcal{W}_f$. As the traction load from the fluid does work on the strip ($\Re(\mathcal{W}_s) >0$), energy is transferred from the former to the latter, with equal and opposite work being done on the fluid ($\Re(\mathcal{W}_f) <0$). To ensure the passivity of the SDoF system, the real part of the work-rate needs to satisfy the condition $\Re(\mathcal{W}_s) \geq 0$. From the admittance of the wall and both Eqs.~(\ref{eq:admittance BC}) and (\ref{eq:work s}), the passivity can be strictly tied to the phase of the admittance wall, $\theta_{Y_{yy}}$, where $\mathcal{W}_s=-|\tilde{v}_s||\tilde{p}| \cos \theta_{Y_{yy}}$. Hence, in order to ensure passivity, the cosine of the admittance phase needs to be either zero or negative. This imposes the requirement that $\theta_{Y_{yy}} \in [\pi/2,3 \pi/2]$, which matches the coupled admittance phase and thus confirms the passivity of the SDoF system. Following a similar process, the passivity of the streamwise admittance, $Y_{xx}$, can be derived from the corresponding traction, $\tilde{t}_{f_{xx}}$, exciting a SDoF system that is reoriented to oscillate in the streamwise direction. Looking at the rate of energy transferred to the strip via the streamwise traction, a similar passivity condition can then be extracted via the admittance phase, yielding a passive streamwise admittance phase in the range $\theta_{xx} \in [0,\pi/2]$ and $[3\pi/2,2\pi]$. With the fluid-structure coupling established, Sec.~\ref{sec:Parametric study} explores the design space of the elastic strip by varying the diagonal terms of the admittance matrix (both amplitude and phase) along with the strip location, with the goal of comprehensively assessing their influence on TS wave growth.

%%%%%%%%%%%%%%%%%%%%%%%%%%%%%%%%%%%%%%%%%%%%%%%%%%%%%%%%
\section{Available design space and analysis of energy routes} \label{sec:Parametric study}

Utilizing the framework established in Sec.~\ref{sec:Comp set}, the interaction of a generalized admittance strip with the TS wave is investigated. The boundary conditions at the admittance strip are given by Eq.~(\ref{eq:admittance BC}). Since the problem of interest is the influence of the strip velocity on the perturbation growth, a prescribed velocity boundary condition is utilized by sweeping across a wide range of velocity amplitudes along with the entire $2 \pi$ phase spectrum. This allows us to compress the design space from eight parameters (phase and amplitude of each admittance term) to four parameters (phase and amplitude of wall-normal and streamwise velocities). The reader is referred to Appendix~\ref{app:VBC} for details regarding the validity of the prescribed velocity boundary conditions. We first examine the effect of the velocity components on the perturbation kinetic energy in the fluid. Following which, a full energy budget analysis is carried out with various strip locations. 

\subsection{Influence of velocity phase and amplitude on TS wave growth} \label{sec:A_P_Sweep}

To understand the mechanism by which the elastic admittance influences TS wave growth, we isolate the response of the strip's velocity components by sweeping each of them independently, i.e., $\tilde{v}_s \neq 0$ while $\tilde{u}_s=0$, and vice versa. The lower and upper bounds of the swept range for each velocity component are given in Table \ref{Tab:BC parameter}. 

\setlength\tabcolsep{1em}
\linespread{1.0}
\begin{table}[htb!]
\centering
\caption{Lower and upper bounds of the swept range of amplitude and phase values for each velocity component in the FEM simulations.}
  \begin{tabular}{cccccc}
    \toprule
    \multirow{2}{*}{} &  Amplitude & units & Phase  & units \\
    %\cmidrule{2-6}
      \midrule
        $\tilde{v}_s$ & $[1\times10^{-7},1\times10^{-13}]$  & [$\frac{\text{m}}{\text{s}}$] & [$0 ,2\pi$] & [rad]\\
		$\tilde{u}_s$ & $[1\times10^{-7},1\times10^{-14}]$ & [$\frac{\text{m}}{\text{s}}$] &  [$0 ,2\pi$] & [rad]\\
    \bottomrule
  \end{tabular}
\label{Tab:BC parameter}
\end{table}

\begin{figure*}[h!]
    \includegraphics[width=1\textwidth]{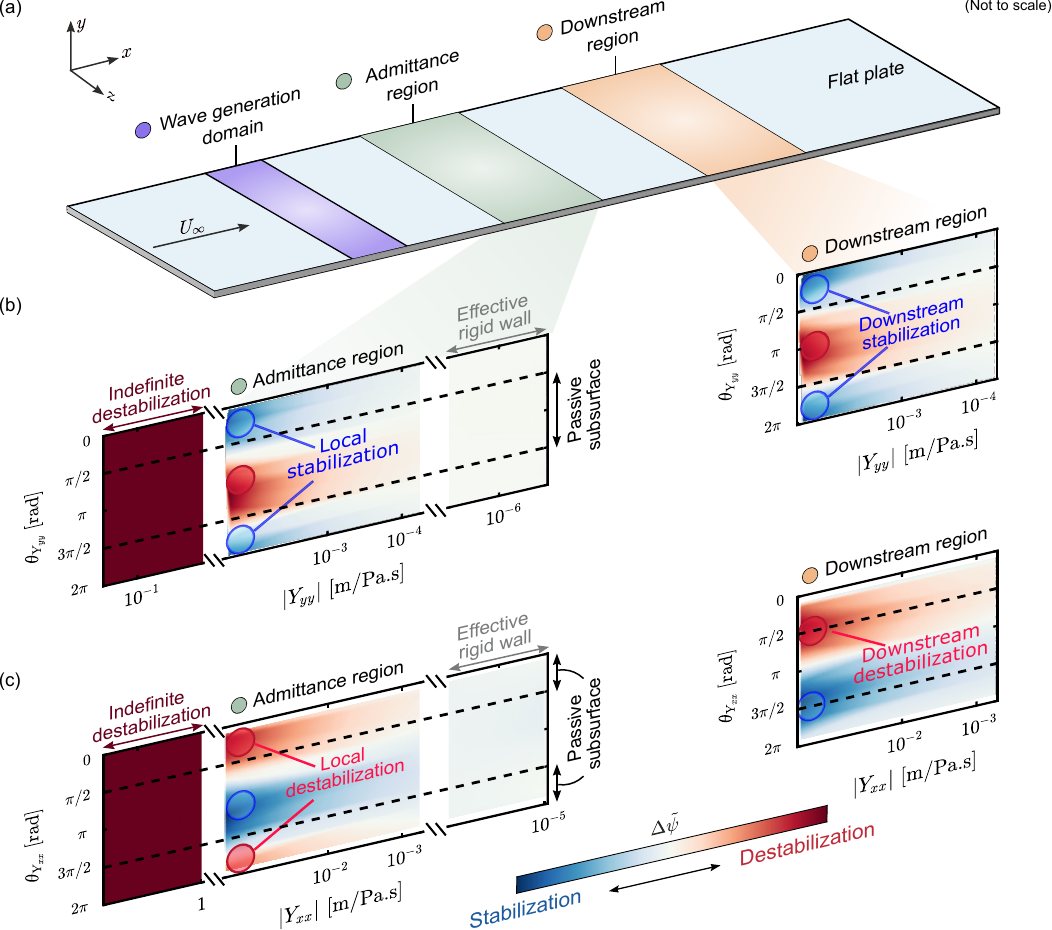}
    \caption{\textbf{Balancing localized and downstream effects of prescribed wall-normal and streamwise velocities on a TS wave instability.} (a) Schematic of a flat plate with free stream velocity, $U_{\infty}$, and three notable regions: Wave generation domain, segment occupied by an elastic strip (modeled as an admittance function), and downstream region encompassing the peak growth of the TS wave. (b) Variations in the normalized contours of the perturbation kinetic energy (PKE), $\Delta \tilde{\psi}$, with changes in amplitude, $|Y_{yy}|$, and phase, $\theta_{Y_{yy}}$, of a wall-normal admittance at the interaction location (left) and downstream of it (right). (c) Same variations with changes in amplitude, $|Y_{xx}|$, and phase, $\theta_{Y_{xx}}$, of a streamwise admittance at the interaction location (left) and downstream of it (right). Regions encompassing extreme admittance amplitudes triggering indefinite TS wave destabilization or no influence on flow (effective rigid wall) are differently shaded. Additionally, phase values physically realizable via a single passive subsurface are also marked for reference.}
    \label{fig:PKE admittance sweep}
\end{figure*}

The extracted perturbation quantities, $\tilde{p}$, $\tilde{\tau}_{xy}$, $\tilde{u}_s$, and $\tilde{v}_s$, at the fluid-structural interface ($y=0$), are averaged along the strip length (i.e., from $x=19.4$ to $19.5$ [cm]). The diagonal admittance terms from Eq.~(\ref{eq:admittance BC}) are extracted from the aforementioned quantities, $Y_{yy}= {\tilde{v}_s}/{\tilde{p}}$ and $Y_{xx} = {\tilde{u}_s}/{\tilde{\tau}_{xy}}$.  The perturbation kinetic energy (PKE), $\tilde{\psi}$, is calculated for the reference (rigid-wall) case and subtracted from each of the swept cases with the fluid interfacing with the admittance strip. The differential PKE is given by
\begin{equation}
\Delta \tilde{\psi} =\frac{1}{4}\rho_{{f}}\int_{0}^{y_{sl}}\left[(|\hat{v}|^2 + |\hat{u}|^2) -(|\hat{v}|^2 + |\hat{u}|^2)_{\mathrm{Ref}}\right]dy 
\label{eq:PKE}
\end{equation}
The derivation of the PKE in the frequency domain is detailed in Appendix~\ref{app:PKE FD}. A negative outcome of Eq.~(\ref{eq:PKE}) indicates a stabilization of the TS waves, i.e., a reduced amount of PKE owing to the strip's interaction with the flow above it. Figure~\ref{fig:PKE admittance sweep}a shows a schematic of the flat plate with a label indicating the wave generation domain exciting the TS waves at $f_e = 500$ [Hz]. Two additional regions are highlighted, one marking the location of the admittance strip and another further downstream. The latter defines the location where the PKE of the reference case reaches its maximum value, which is a function of the speed, frequency, and the selected amplification rate from the neutral curves. The differential PKE is averaged along both regions and is plotted (normalized by the maximum value) versus $|Y_{yy}|$ and $\theta_{Y_{yy}}$ in Fig.~\ref{fig:PKE admittance sweep}b. Details on how $\theta_{Y_{yy}}$ is computed are provided in Appendix \ref{app:phase_comp}. The plots show the effect of the wall-normal velocity boundary condition on the PKE. The left panel of the figure shows the ``local'' influence of the wall-normal admittance and phase on the TS wave behavior, i.e., at the immediate location of the flow's interaction with the elastic strip. Looking separately at the amplitude effect, two important observations can be made. First, cut-off thresholds can be identified at small amplitudes where the strip has negligible effect on the growth of the TS waves, effectively rendering the strip a rigid wall at the regions labeled as such on the figure. More importantly, another threshold ($|Y_{yy}| \approx  10^{-1}$ [m/Pa$\cdot$s]) can be identified, as labeled on the figure, above which the admittance strip is indefinitely destabilizing for all phase values. 

Excluding these two extreme ends of the spectrum where the strip is either too rigid to impose a tangible effect on the flow or too soft that its effect on the TS wave is always detrimental, we can now focus our analysis on the intermediate $|Y_{yy}|$ values in between. Within this intermediate range, while the color contours remain fairly consistent across changes in admittance amplitude, the threshold marking the transition between stabilization and destabilization (white contours in Fig.~\ref{fig:PKE admittance sweep}) exhibits small variations in the phase values with changes in amplitude. Accordingly, the phase intervals identified below should be interpreted as approximate ranges. That said, examining the phase effect on TS wave growth across these intermediate admittance amplitudes remains of critical importance. Here, it is worth reminding that the admittance phase for a passive elastic strip was shown in Sec.~{\ref{sec:FSI}} to fall in the range defined by $\theta_{Y_{yy}} \in [\pi/2 ,3 \pi/2]$\footnote{Note that the same range for an uncoupled strip (i.e., from a pure structural vibrations analysis) lies within the $[-\pi/2,\pi/2]$ range crossing the zero phase at resonance. This shift in the range is due to the retaining of a negative sign in the fluid traction acting on the structure, $-\tilde{p}$, when the strip is fully coupled with the flow. \label{FN:phase_shift}}. Figure~\ref{fig:PKE admittance sweep}b reveals that $\theta_{Y_{yy}}$ falling between $7\pi/12$ and $19\pi/12$ induces a destabilizing effect on the TS wave both locally over the strip and downstream of it, which is doubly undesirable. However, three phase windows are shown to achieve the desirable combination of local and downstream stabilization, i.e., complete attenuation of the TS wave along the entire streamwise direction. The first is for $\theta_{Y_{yy}}$ values falling between $\pi/2$ and $7\pi/12$. While this narrow window falls within the range obtainable by a single passive admittance strip, the current results are obtained by imposing a wall-normal admittance as a modeling technique rather than an actual subsurface structure, allowing us to purely impose a wall-normal velocity ($\tilde{v}_s \neq 0$ while $\tilde{u}_s$ remains $0$), as explained in Sec.~\ref{sec:FSI}. The inherent coupling between the velocity components renders the above result unachievable in practice. Realistically, a subsurface structure which vibrates in the $y$-direction practically displaces ``into'' the fluid causing an oscillatory, albeit small, bump in the wall as a result of the elastic deformation. To account for this condition, a modification is made by linearizing the flow around the bump, resulting in a streamwise velocity boundary condition ($\tilde{u}_s \neq 0$) coming purely from the wall-normal displacement. The second and third favorable phase windows are for $\theta_{Y_{yy}}$ values falling between $0$ and $\pi/2$, and between $19\pi/12$ and $2\pi$. Nevertheless, the passivity condition established in Sec.~\ref{sec:FSI} is not met for either range, rendering the results also unachievable by using a passive structure which interfaces with the flow at a single location. The previous analysis explains, in a canonical manner, why the highly favorable outcome of stabilizing the TS wave both locally and downstream via a single subsurface structure has been thus far elusive.

\newpage
Figure~\ref{fig:PKE admittance sweep}c sheds light on the effect of imposing a streamwise admittance, $Y_{xx}$, on the behavior of the TS wave, when a velocity boundary condition of the form $\tilde{u}_s \neq 0$ and $\tilde{v}_s =0$ is imposed. The two plots demonstrate the effects of both the amplitude and phase of such admittance, $|Y_{xx}|$ and $\theta_{Y_{xx}}$, respectively, on the PKE at the location of the admittance and further downstream. Similar to Fig.~\ref{fig:PKE admittance sweep}b, values of $|Y_{xx}| > 1$ [m/Pa$\cdot$s] correspond to indefinite destabilization while $|Y_{xx}| < 10^{-5}$ [m/Pa$\cdot$s] results in an effective rigid wall. However, two interesting differences between Figs.~\ref{fig:PKE admittance sweep}b and c arise which highlight the fundamental differences between how wall-normal and streamwise admittances affect the flow perturbation. The first pertains to the amplitude. The streamwise admittance is able to achieve negative values of $\Delta \tilde{\psi}$ (i.e., stabilization relative to the reference rigid-wall scenario) at one order of amplitude less than the wall-normal admittance, doing away with the need for an aggressively soft material to provide the necessary control mechanism. The second notable difference is the opposite effect of the phase on the TS wave attenuation. A strip oscillating strictly in the streamwise direction is able to attenuate the TS waves locally when $\theta_{Y_{xx}}$ falls between $3\pi/4$ and $7\pi/4$, but not necessarily sustain downstream stabilization. On the other hand, $\theta_{Y_{xx}}$ between $\pi$ and $2\pi$ provides a favorable downstream effect. The intersection of these two ranges yields a window of $\theta_{Y_{xx}}$ values between $\pi$ and $7\pi/4$ which appears to achieve both local and downstream stabilization. However, only the higher end of this range, i.e., $3\pi/2<\theta_{Y_{xx}}<7\pi/4$, also satisfies the strip's passivity criterion, providing evidence that a single passive strip could be used to achieve this optimal outcome without the need for additional interventions. This new avenue for sustained TS wave attenuation will be explored in more detail in Secs.~\ref{sec:SDR} and \ref{sec:physical realization of streamwise adm}.

%%%%%%%%%%%%%%%%%%%%%%%%%%%%%%%%%%%%%%%%%%%%%%%%%%%%%%%%%
\subsection{Energy budget analysis}\label{sec:EB}

We conduct a full energy budget analysis to further understand the effect of the phase on the TS wave growth. Starting from the LNS equations, shown in Eq.~(\ref{eq:LNS}), with a portion of the flat plate replaced with the admittance strip, and upon proper treatment (see Appendix \ref{app:Energy_budget} for details), the following energy budget equation is obtained \cite{Morris1976},
\begin{equation}
\begin{aligned}
\frac{d}{dx}
\left[
\overbrace{\int_0^{y_\infty} U{{\tilde{\psi}}_k}\,dy}^{(\text{I})}
+\overbrace{\int_0^{y_\infty}\frac{1}{\rho_{{f}}} \tilde u\tilde p\,dy}^{(\text{II})}
-\overbrace{\int_0^{y_\infty}\frac{\mu_{{f}}}{\rho_{{f}}} \tilde v\tilde \xi\,dy}^{(\text{III})}
\right]
=
\underbrace{-\int_0^{y_\infty}\tilde u\tilde v\,U_y\,dy}_{(\mathcal{P})}
-
\\[4pt]
\underbrace{\int_0^{y_\infty} \frac{\mu_{{f}}}{\rho_{{f}}}\tilde \xi^2\,dy}_{(\mathcal{D})}
-\underbrace{\int_0^{y_\infty}\tilde u\tilde v\,V_x\,dy}_{(\text{i})}
+\underbrace{\int_0^{y_\infty}
(\tilde v^2-\tilde u^2)\,U_x\,dy}_{(\text{ii})}
+\underbrace{\left. \frac{1}{\rho_{f}}\tilde v_s\tilde p \right|_{y=0}}_{(\mathcal{W})}
+\underbrace{\left. \frac{\mu_{{f}}}{\rho_{{f}}} \tilde u_s\tilde \xi\right|_{y=0}}_{(\Lambda)}
\end{aligned}
\label{eq:Energy Budget}
\end{equation}
where $\tilde{\psi}_k=\frac{1}{2}\left(\left|\tilde{u}\right|^2 + \left|\tilde{v}\right|^2\right)$ is the pointwise specific PKE, and $\tilde{\xi}=\tilde v_x-\tilde u_y$ is the perturbation vorticity. In Eq.~(\ref{eq:Energy Budget}), $(.)_{x}$ and $(.)_{y}$ denote the partial derivatives with respect to the streamwise and wall-normal coordinates, respectively, while $y_\infty$ denotes the far-field location in the wall-normal direction, taken here to coincide with $y_{sl}$. The equation is formulated such that the left-hand side represents the rate of change of the integrated perturbed energy flux, consisting of three contributions: The mean-flow convection of perturbed kinetic energy (term~I), the pressure-related transport of disturbance energy (term~II), and the viscous transport contribution (term~III). These terms, when combined, characterize the advection and redistribution of disturbance energy across a given streamwise cross section. It is evident that (term~III) is negligible compared to the contributions of the other terms \cite{Davies1997}.

On the right-hand side, $\mathcal{P}$ denotes the production term arising from the interaction between the Reynolds shear stress, $\tilde{u}\tilde{v}$, and the wall-normal gradient of the streamwise base flow, $U_y$, thereby quantifying the transfer of energy from the mean flow to the perturbation. $\mathcal{D}$ is the viscous dissipation term which accounts for the perturbation energy sink, in contrast to the production term, which acts as a source. An additional production mechanism (term i) is associated with the streamwise gradient of the wall-normal component of the base-flow velocity, whereas the remaining production contribution (term~ii) is associated with the Reynolds normal stresses, $\tilde{u}^2$ and $\tilde{v}^2$. Finally, the remaining terms denote the admittance's wall-contribution terms: $\mathcal{W}$ represents the rate of pressure work done by the structure on the fluid (with $\mathcal{W} = \mathcal{W}_f/\rho_f$) and $\Lambda$ corresponds to the viscous energy at the wall. 

For our application of flow over a flat plate, these contributions are dominated by the production and dissipation terms, $\mathcal{P}$ and $\mathcal{D}$, in addition to the interfacial terms, $\mathcal{W}$ and $\Lambda$. The remaining terms are expected to be small for the slowly developing boundary layer considered here and are neglected in the subsequent analysis. Moreover, rather than considering the perturbation-energy flux on the left-hand side of Eq.~(\ref{eq:Energy Budget}), the evolution of the PKE is examined directly since its spatial growth (or decay) is one of the most commonly used measures in literature to assess the destabilizing (or stabilizing) effect of the elastic strip. The PKE is, therefore, used to quantify the instability response, while variations in the dominant energy-budget right-hand-side terms are used to elucidate the mechanisms associated with each of the two principal boundary conditions given by Eq.~(\ref{eq:admittance BC}), under variations in phase and streamwise location.

\begin{figure}[htbp]
    \centerline{\includegraphics[width=1\columnwidth]{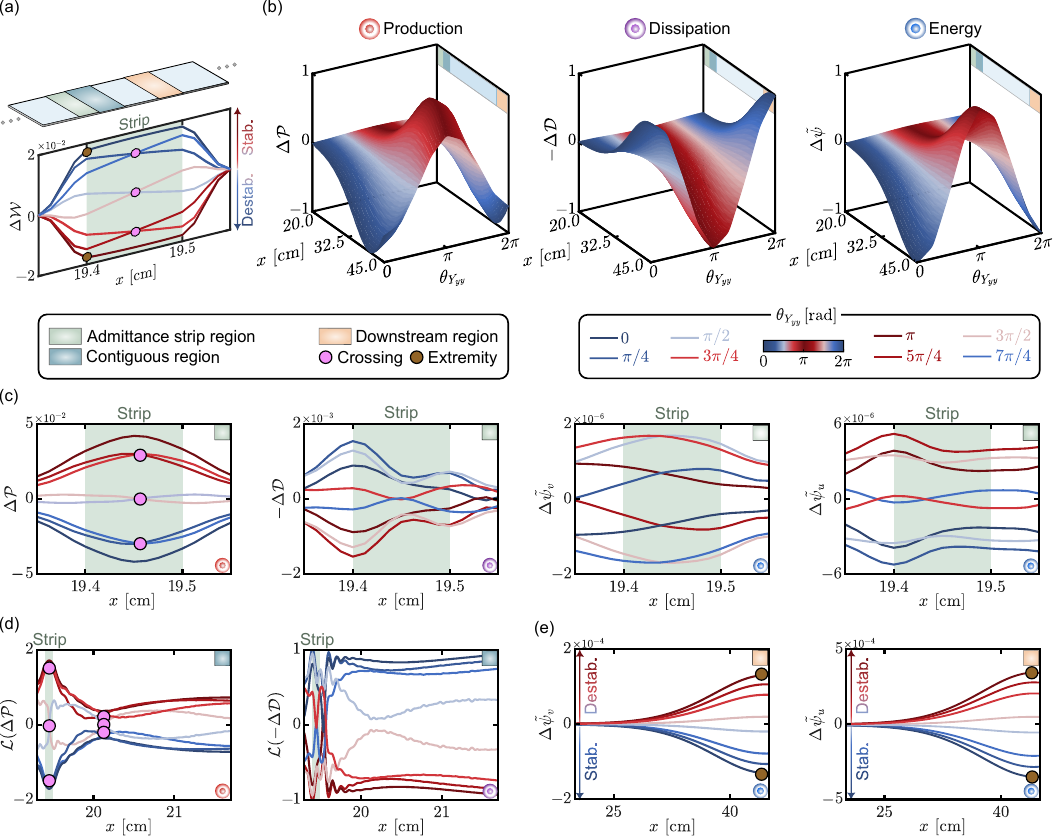}}
\caption{\textbf{Effect of relative phasing between the prescribed wall-normal velocity phase and the pressure, captured by the wall-normal admittance phase, $\theta_{Y_{yy}}$, on TS wave growth.} (a) Schematic depicting the strip, contiguous, and downstream regions (top), and spatial variation of the interfacial work-rate term, $\Delta \mathcal{W}$, for the eight listed and color-coded $\theta_{Y_{yy}}$ values (bottom). (b) Variations relative to the reference case in production, $\Delta \mathcal{P}$ (left), dissipation, $-\Delta \mathcal{D}$ (middle), and PKE, $\Delta \tilde{\psi}$ (right), along the streamwise direction and over the full range of $\theta_{Y_{yy}}$. (c) Spatial profiles of the changes relative to the reference case in the production, $\Delta\mathcal{P}$ (left), dissipation, $-\Delta\mathcal{D}$ (middle left), wall-normal PKE, $\Delta\tilde{\psi}_{v}$ (middle right), and streamwise PKE, $\Delta\tilde{\psi}_{u}$ (right), within the strip region. (d) Spatial profiles of the changes compared to the reference case in the production, $\mathcal{L}(\Delta\mathcal{P})$ (left), and dissipation, $\mathcal{L}(-\Delta\mathcal{D})$ (right), throughout the contiguous region represented using a tailored logarithmic scale. (e) Evolution of the wall-normal (left) and streamwise (right) PKE variations, $\Delta\tilde{\psi}_{v}$ and $\Delta\tilde{\psi}_{u}$, starting from the end of the contiguous region into the \mbox{downstream region.}}
    \label{fig:vsweep}
\end{figure}

\subsubsection{Prescribed wall-normal velocity phase \label{Sec:W_P}}

In this analysis, the growth of the TS wave mode is examined over three distinct regions. The first two are the same regions defined in Sec.~\ref{sec:A_P_Sweep}, namely, the admittance strip location and a downstream location corresponding to the peak growth of the instability of the reference (rigid-wall) case. The third is a contiguous region sitting between the aforementioned two, where a transition effect takes place, which will be discussed later. Figure~\ref{fig:vsweep}a\footnote{We note that in all of Figs.~\ref{fig:vsweep} and \ref{fig:usweep} (except for part b in both), the plotted quantities are normalized by a constant numerical value equal to the peak PKE of the rigid-wall case for optimal data visualization.} depicts the difference in the work-rate, $\Delta \mathcal{W} = \mathcal{W} - \mathcal{W}_{\mathrm{Ref}}$, which best captures the interaction between the fluid and structural domains (We note that $\Delta \mathcal{W}$ is effectively equal to the work-rate of the strip since the work-rate of the reference case, $\mathcal{W}_{\mathrm{Ref}}$, is zero). Accordingly, eight cases with constant admittance amplitude, $|Y_{yy}| = 1.725 \times 10^{-5}$ [m/Pa$\cdot$s], and uniformly spaced phase differences, $\theta_{Y_{yy}}$, are used to assess the influence of this term. The in-phase ($\theta_{Y_{yy}} =0$) and out-of-phase ($\theta_{Y_{yy}} =\pi$) cases produce the largest absolute response in the wall work-rate with positive and negative signs, respectively, as shown in the figure, while the quadrature cases ($\theta_{Y_{yy}} =\pi/2$ and $\theta_{Y_{yy}} =3\pi/2$) produce a response which is  close to zero. The remaining cases produce a behavior which lies in between. Additionally, unlike the rest of the $\theta_{Y_{yy}}$ values, $\mathcal{W}$ appears to be completely flat over the interval housing the strip for both the in-phase and out-of-phase cases. This behavior can be attributed to the small variation of the perturbation pressure along the strip (Note that $\mathcal{W} \sim |\tilde{v}_s||\tilde{p}|\cos{\theta_{Y_{yy}}}$ at $y=0$). For this two specific phase values, the cosine function has a zero slope (local extrema). Any variation around the local extrema results in a small perturbation in phase while the sign remains unchanged, resulting in the near-flat behavior. On the other hand, a small variation in the perturbation pressure in the remaining cases yields a considerable variation in the work-rate term due to the large slope of the cosine function at these phase angles. 

Figure~\ref{fig:vsweep}b presents the variation of the PKE (right) and the two main mechanisms, production (left) and dissipation (middle), relative to the reference scenario, across the three regions and over the full span of phase values, with each quantity normalized by its maximum absolute value. Similar to Fig.~\ref{fig:vsweep}a, these surface plots show the difference, $\Delta$, between each respective quantity and its counterpart in the reference case. Hence, positive values indicate an increase with respect to the reference case, whereas negative values indicate a decrease. Furthermore, dissipation is shown through its negative contribution, $-\Delta \mathcal{D}$, consistent with the energy-budget form of Eq.~(\ref{eq:Energy Budget}). The three plots show almost the same behavior (flipped for dissipation) throughout the downstream region, despite the different local behaviors at the strip location (which can be best seen in the close-ups of Fig.~\ref{fig:vsweep}c). Moreover, the behavior follows, to a great extent, the amplitude response of the work-rate term, indicating a direct link between these quantities. The same phenomenon indicated earlier is observed as a function of phase, such that the maximum absolute values occur at $\theta_{Y_{yy}} =0$ and $\theta_{Y_{yy}} =\pi$, regardless of the quantity.

Figure~\ref{fig:vsweep}c provides a closer look at $\Delta \mathcal{P}$, $-\Delta \mathcal{D}$, and the PKE, which is is decomposed into the streamwise and wall-normal components, $\Delta\tilde{\psi}_{v}$ and $\Delta\tilde{\psi}_{u}$, versus the streamwise coordinate $x$ in the vicinity of the strip location (which starts at $x_s = 19.4$ [cm]). The work-rate done by the strip contributes to an explicit change in the wall-normal PKE, $\Delta\tilde{\psi}_{v}$, as shown by the right most panel of the figure. At first glance of Eq.~(\ref{eq:PKE}), it may be counterintuitive to observe variations in the wall-normal PKE at the same admittance amplitude. However, the variation in phase of the prescribed velocity at the wall modulates the mode shape of the wall-normal velocity which is integrated across the domain height to obtain the wall-normal PKE component. Therefore, local stabilization or destabilization of the wall-normal PKE is governed mainly by the phase relationship between the reference case and the impact of the prescribed velocity boundary condition along the wall-normal direction throughout the boundary layer thickness. Although the energy variations of the wall-normal velocity component appear to be uncorrelated with the rate of work, they are implicitly linked to it. Despite the strong near-wall response due to the admittance strip, the wall-normal velocity mode shape experiences phase variation along the wall-normal direction. As a result, the wall-normal PKE, which is integrated over the wall-normal direction, differs from the work-rate response. That said, four of the eight cases shown in Fig.~\ref{fig:vsweep}c exhibit local stabilization of the wall-normal PKE, $\Delta\tilde{\psi}_{v}$, whereas the remaining cases undergo destabilization, indicating that the shift in the speculated TS wave attenuation or amplification behavior is consistent across all cases.

The inherent coupling between the streamwise and wall-normal velocity components also contributes to the local response of the streamwise velocity and its associated PKE, $\Delta\tilde{\psi}_{u}$. By comparing the two PKE components, $\Delta\tilde{\psi}_{u}$ and $\Delta\tilde{\psi}_{v}$, at the strip location, it can be seen that an in-phase ($\theta_{Y_{yy}} =0$) scenario triggers local attenuation in both PKE components, while an out-of-phase ($\theta_{Y_{yy}} =\pi$) scenario induces local amplification in both PKE components, as evident by their entirely-negative values over the plotted length for the former case and their entirely-positive values for the latter case. The two PKE components, however, behave oppositely in the quadrature cases. Additionally, we observe that phase lags and leads of $\pi/4$ relative to the in-phase and out-of-phase conditions produce the minimum and maximum absolute variations in the streamwise PKE, $\Delta\tilde{\psi}_{u}$, respectively. Since the overall PKE is a summation of both components, the strongest localized PKE attenuation relative to the reference case is obtained for phase differences of $\pi/4$ followed by $0$, whereas the strongest localized PKE amplification is obtained for phase differences of $5\pi/4$ followed by $\pi$. 

The impact of work-rate is not limited to the perturbed velocity (PKE) components, it also extends to the main governing mechanisms: Production and dissipation. These two terms become important when downstream control of the TS wave is targeted. At a sufficient distance from the strip location, the change of perturbed energy flux given by the left-hand-side terms of Eq.~(\ref{eq:Energy Budget}), and consequently the PKE behavior, is governed by the balance between production and dissipation. Hence, we track both terms from the strip location to the location at which the PKE peaks. Initially, production opposes the work-rate contribution of the strip, as demonstrated by $\Delta \mathcal{P}$ in the most left panel of Fig.~\ref{fig:vsweep}c. When energy is transferred from the fluid to the structure through a negative rate of work, $\mathcal{W}=-\mathcal{W}_s/\rho_f$ (red cases in Fig.~\ref{fig:vsweep}a and c), it strengthens the correlation between the two velocity components, represented by the Reynolds shear stress, whereas adding energy to the fluid weakens this correlation. Meanwhile, dissipation largely follows the variation in the streamwise PKE, but with an opposite sign, as evident by comparing the $-\Delta\mathcal{D}$ and $\Delta\tilde{\psi}_{u}$ plots. This resemblance arises because the vorticity depends heavily on the wall-normal gradient of the streamwise velocity.

Moving from the strip location to the adjacent contiguous region, the field attempts to recover from the disturbance imposed by the strip, as shown in Fig.~\ref{fig:vsweep}d. During this recovery process, production tends toward the reference state for all phase values, as shown in the leftmost panel of the figure\footnote{For better visualization, a signed logarithmic operator is adopted outside the immediate vicinity of the strip. For a plotted quantity $q$, the transformed value $\mathcal{L}(q)$ is defined as $\mathcal{L}(q) = \operatorname{sgn}(q)\, \log_{10}\!\big(1+\frac{|q|}{q_0}\big)$, where $q_0$ is a reference scale used to set the size of the linear region near zero. This transformation preserves the sign of the original quantity, compresses large positive and negative values logarithmically, and avoids the singularity of a standard logarithm at $q=0$.}. However, before reaching zero, most of the curves exhibit a rebound. The quadrature cases form an the exception, as they cross the zero line and continue growing on the opposite side. Dissipation also undergoes a recovery process, albeit over a shorter distance, as shown in the second panel of Fig.~\ref{fig:vsweep}d.
Further downstream, all curves eventually follow the conventional growth trend predicted by the LST. In the final state, the relative ordering of the dissipation curves with respect to one another appears to mirror the production term about the zero axis. Because the downstream region is governed mainly by these two mechanisms, both PKE components, $\Delta\tilde{\psi}_{u}$ and $\Delta\tilde{\psi}_{v}$, shown in Fig.~\ref{fig:vsweep}e, exhibit an arrangement that, at sufficiently large distances from the strip, reflects the final configurations of the production and the mirrored dissipation observed earlier in the surface plots of Fig.~\ref{fig:vsweep}b. This final configuration of the eight cases is also approximately the mirrored response of the work-rate term. The in-phase and out-of-phase cases achieve the strongest manipulation, whereas the cases that have a crossing point in the interface term, highlighted by the pink circular markers in Fig.~\ref{fig:vsweep}a, are arranged based on the amplitude of the work-rate response at the left side of the strip.

\subsubsection{Prescribed streamwise velocity phase\label{Sec:S_P}}

\begin{figure}[htbp]
\centerline{\includegraphics[width=1\columnwidth]{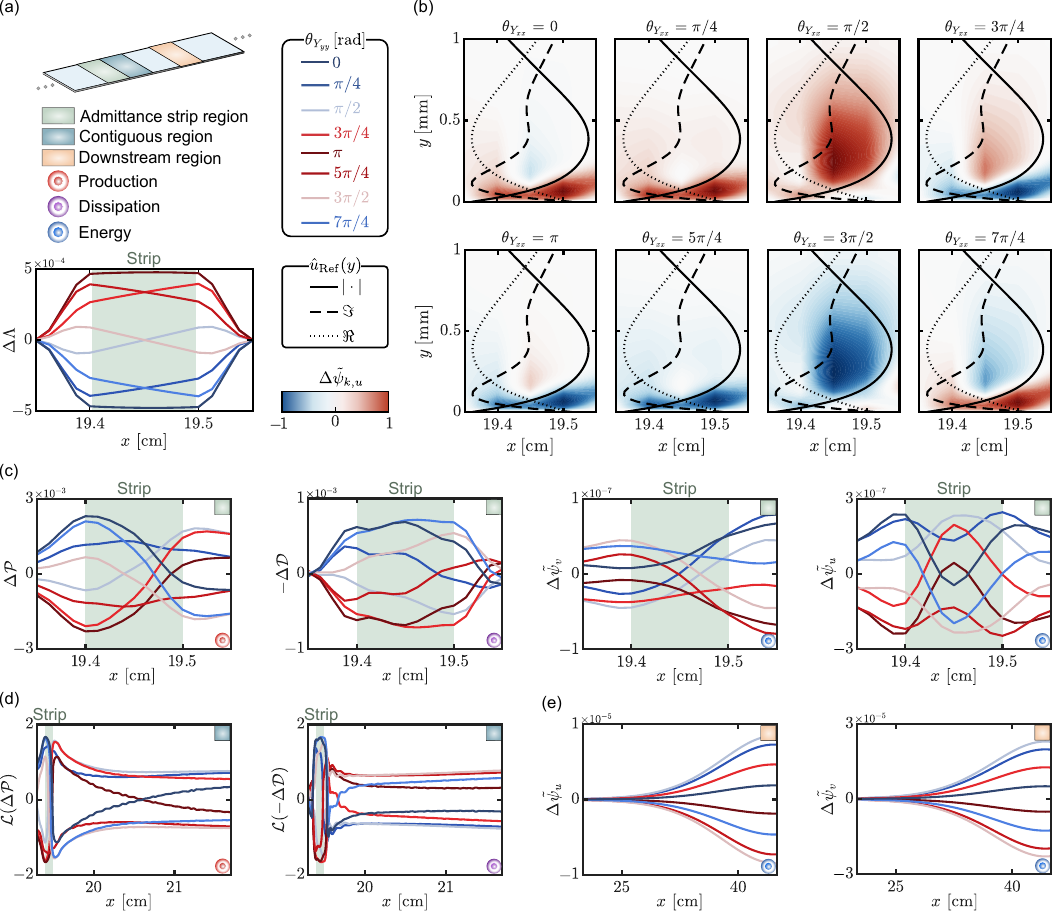
}}
\caption{
\textbf{Effect of relative phasing between the prescribed streamwise velocity and the shear stress, captured by the streamwise admittance phase, $\theta_{Y_{xx}}$, on TS wave growth.} (a) Schematic depicting the strip, contiguous, and downstream regions (top), and spatial variation of the interfacial viscous energy term, $\Delta \Lambda$, for the eight listed and color-coded $\theta_{Y_{xx}}$ values (bottom). (b) Pointwise variations of the streamwise PKE, $\Delta\tilde{\psi}_{k,u}$, within the strip region, in addition to the real part, imaginary part, and magnitude of the streamwise-velocity mode shape of the reference (rigid-wall) case. (c) Spatial profiles of the changes relative to the reference case in the production, $\Delta\mathcal{P}$ (left), dissipation, $-\Delta\mathcal{D}$ (middle left), wall-normal PKE, $\Delta\tilde{\psi}_{v}$ (middle right), and streamwise PKE, $\Delta\tilde{\psi}_{u}$ (right), within the strip region. (d) Spatial profiles of the changes compared to the reference case in the production, $\mathcal{L}(\Delta\mathcal{P})$ (left), and dissipation, $\mathcal{L}(-\Delta\mathcal{D})$ (right), throughout the contiguous region represented using a tailored logarithmic scale. (e) Evolution of the wall-normal (left) and streamwise (right) PKE variations, $\Delta\tilde{\psi}_{v}$ and $\Delta\tilde{\psi}_{u}$, starting from the end of the contiguous region into the downstream region.}
\label{fig:usweep}
\end{figure}

The second key ingredient in controlling the TS wave is the streamwise velocity boundary condition, whose primary effect appears as a variation in the viscous term, $\Delta \Lambda=\Lambda-\Lambda_{\mathrm{Ref}}$, as shown in Fig.~\ref{fig:usweep}a, when using a constant streamwise admittance amplitude of $|Y_{xx}| = 6.556 \times 10^{-4}$ [m/Pa$\cdot$s]. Once again, we note that $\Lambda_{\mathrm{Ref}}=0$ for the reference case, and therefore $\Delta \Lambda=\Lambda$. Although this term represents a coupling between the vorticity and the streamwise velocity, the vorticity can be approximated as $\tilde{\xi}\approx-\tilde{u}_y\approx-\tilde{\tau}_{xy}/\mu_f$, because of the dominance of the wall-normal gradient of the streamwise velocity, $\tilde{u}_y$, over the streamwise gradient of the wall-normal velocity, $\tilde{v}_x$, particularly at the strip. Thus, similar to $\mathcal{W}$, $\Lambda$ can effectively be considered as a coupling between the streamwise traction (shear stress) and the streamwise velocity. As a result, a streamwise admittance phase of $\theta_{Y_{xx}}=0$ (in-phase) or $\theta_{Y_{xx}}=\pi$ (out-of-phase) produces a nearly flat local response, whereas all other phases values yield spatial variation over the strip region, similar to the $\mathcal{W}$ behavior previously observed in Fig.~\ref{fig:vsweep}a.

Locally, at the strip region, the streamwise PKE, $\Delta\tilde{\psi}_u$, exhibits oscillatory spatial variation, as captured by the rightmost panel of Fig.~\ref{fig:usweep}c. This behavior likely arises from scattering caused by the abrupt change in the boundary condition at the strip edges, together with the phase variation of the reference streamwise velocity, $\tilde{u}_{\mathrm{Ref}}$, along the wall-normal direction and the fact that the peak of its mode shape, $\hat{u}_{\mathrm{Ref}}(y)$, is located close to the wall. As a result, the modulation of the streamwise-pointwise PKE, $\Delta \tilde{\psi}_{k,u}=\frac{1}{2}\Delta \tilde{u}^2$, normalized by its maximum absolute value, shown in Fig.~\ref{fig:usweep}b, reveals two distinct layers of variation: A near-wall region characterized by left and right hot spots, and a high-intensity region located near the peak of the mode shape.

To elaborate on how the phase change of the streamwise velocity in the wall-normal direction together with wall-normal location of the mode shape contributes to the oscillatory behavior, the two components, $2\tilde{u}_{\mathrm{Ref}}\delta_u$ and $\delta_u^2$, resulting from the expansion of the squared perturbed streamwise velocity, $\tilde{u}^2=(\tilde{u}_{\mathrm{Ref}}+\delta_u)^2$, relative to its reference counterpart, are examined, where $\delta_u$ denotes the spatial disturbance induced by the prescribed streamwise velocity. At the vicinity of the wall, where $\tilde{u}_{\mathrm{Ref}}$ tends to zero, the $2\tilde{u}_{\mathrm{Ref}}\delta_u$ term dominates and therefore explains the observed spatial pattern. At the wall, the pointwise PKE variation is approximately zero because $\tilde{u}_{\mathrm{Ref}}=0$ and $\delta_u$ is relatively small. However, the reference mode shape, $\hat{u}_{\mathrm{Ref}}(y)$, grows rapidly away from the wall, resulting in a significant increase in the pointwise PKE variation. Moreover, the near-wall region becomes increasingly influenced by $\tilde{u}_{\mathrm{Ref}}$ which approximately retains the phase signature of the reference perturbed wall shear stress (at $y=0$), while $\delta_u$ still carries the phase signature of the prescribed velocity. In other words, the phase difference between the reference case streamwise velocity and the prescribed velocity disturbance within the near-wall region remains similar to that between the wall shear and the prescribed velocity at the wall. As a result, the response in this region is primarily governed by the streamwise admittance phase, $\theta_{Y_{xx}}$, i.e., the phase between the imposed streamwise velocity and the streamwise traction, represented by the shear stress.

Further from the wall, the direct influence of the boundary condition weakens and the response becomes increasingly distorted. Nevertheless, looking at Fig.~\ref{fig:usweep}b, by the time the reference mode shape, $\hat{u}_{\mathrm{Ref}}(y)$, reaches its maximum magnitude, it also undergoes a significant phase shift (evident by the profiles of the real, $\Re$, and imaginary, $\Im$, components), producing a different higher (i.e., away from the wall) high-intensity spot than its lower (near-wall) counterpart. The contour plots also indicate that quadrature streamwise admittance phase values, $\theta_{Y_{xx}}=\pi/2$ and $3\pi/2$, lead to the weakest lower spots, compared to the remaining cases which strengthen the response near the wall and extend this lower spot region slightly upwards. Although the wall-normal PKE exhibits variations that are not aligned with those of the streamwise component, as shown in Fig.~\ref{fig:usweep}c, the trend of the total PKE, $\Delta \tilde{\psi}$, still approximately follows that of the streamwise PKE component, $\Delta \tilde{\psi}_u$. 

Similar to the prescribed wall-normal velocity boundary condition, the way that the PKE responds downstream originates from local variations in production and dissipation at the strip region, as shown in Fig.~\ref{fig:usweep}c. However, the production behavior within a strip with a streamwise admittance differs markedly from that observed in Sec.~\ref{Sec:W_P}. Specifically, six of the eight $\Delta \mathcal{P}$ cases shown in the leftmost plot of Fig.~\ref{fig:usweep}c exhibit a zero crossing. The $\theta_{Y_{xx}}=\pi/2$ and $\theta_{Y_{xx}}=\pi$ cases increase from negative to positive values, contrary to the trend observed for $\theta_{Y_{xx}}=0$ and $\theta_{Y_{xx}}=3\pi/2$. The intermediate phase angles, $\theta_{Y_{xx}}=3\pi/4$ and $\theta_{Y_{xx}}=7\pi/4$, transition from an out-of-phase and in-phase behavior on the left side of the strip to a quadrature behavior on the right, respectively, while the $\theta_{Y_{xx}}=\pi/4$ and $\theta_{Y_{xx}}=5\pi/4$ cases remain strictly positive or negative. Although the local production response does not directly follow the viscous term, every case that exhibits a progressive increase in $\Delta \Lambda$ from the left to the right side of the strip ends up with a positive production sign ($\Delta \mathcal{P} >0$) at the right edge of the strip ($x = 19.5$ [cm]). Additionally, even the out-of-phase case, which exhibits a flat production rather than a progressive increase, crosses the zero into positive territory shorty before the strip ends. In contrast, the $-\Delta \mathcal{D}$ dissipation curves exhibit a zero crossing only for the cases that do not cross zero in the production window, i.e., $\theta_{Y_{xx}}=\pi/4$ and $\theta_{Y_{xx}}=5\pi/4$, thereby, in a broader sense, following the behavior of the $\Delta \Lambda$ term.

Post strip, the production and dissipation mechanisms begin to recover, but none of the cases fully returns to its original state. In the left plot of Fig.~\ref{fig:usweep}d, only two cases cross the zero axis and settle on the opposite side; a pattern which is somewhat similar to that observed with a wall-normal admittance in Sec.~\ref{Sec:W_P}. In the latter, however, the two zero crossings corresponded to the quadrature phase values, whereas with the current streamwise admittance, they correspond to the in-phase and out-of-phase configurations. On the dissipation side, shown in the right plot of Fig.~\ref{fig:usweep}d, the recovery exhibits a different pattern, but the final arrangement yields a flipped version of the production and the two PKE components, shown in \ref{fig:usweep}e.

\begin{figure}[htbp]
    \centerline{\includegraphics[width=\columnwidth]{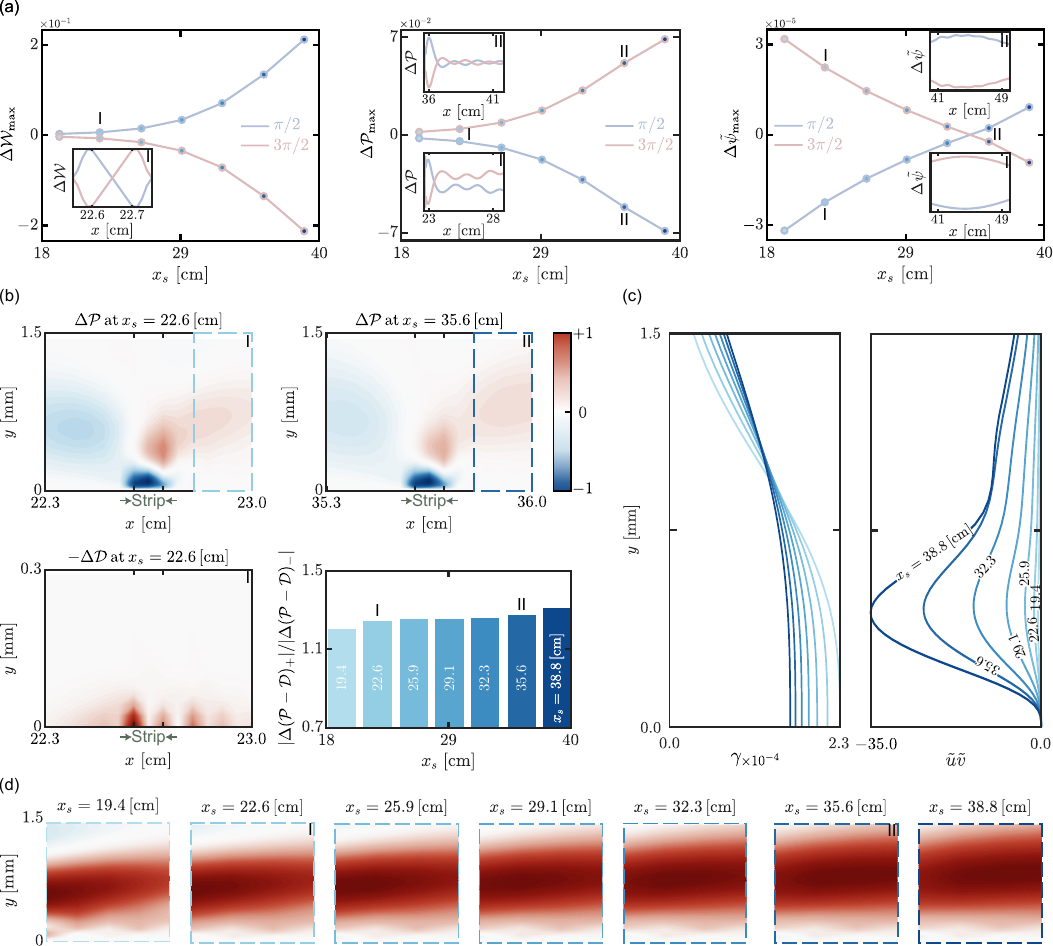}}
    \caption{\textbf{Influence of strip location on TS wave growth in response to a prescribed wall-normal velocity.} (a) Effect of changing the strip location, $x_s$, on the maximum signed values of the interfacial work-rate, $\Delta {\mathcal{W}}_{\max}$ (left), production, $\Delta {\mathcal{P}}_{\max}$ (middle), and the downstream PKE, $\Delta {\tilde \psi}_{\max}$ (right). The small insets show the spatial profiles of each of the three aforementioned parameters, for select $x_s$ values, from which the maximum values are extracted. (b) Production levels compared to the reference case, $\Delta \mathcal{P}$ (top), corresponding to $x_s=22.6$ [cm] (left) and $x_s=35.6$ [cm] (right), and dissipation levels compared to the reference case, $-\Delta \mathcal{D}$, corresponding to $x_s=22.6$ [cm] (bottom left). The bottom right panel shows the effect of strip location on the positive-to-negative extrema ratio of the vertically integrated combined production and dissipation variations, $\Delta(\mathcal{P}-\mathcal{D})$. (c) The variation along the wall-normal direction, $y$, in the shear rate profile, $\gamma$ (left), and the corresponding Reynolds shear stress (right). (d) Close-up insets of the region bounded in part (b), showing the strengthening of the extended tail of the positive production pole for all seven strip locations, $x_s$, studied here.}
\label{fig:vlocation}
\end{figure}

\subsubsection{Prescribed wall-normal velocity location \label{sec:W_L}}

The effect of the admittance strip location on the growth of the TS instability is investigated from an energy-budget perspective. Using the same strip length ($\ell_s=1$ [mm]) and wall-normal admittance amplitude ($|Y_{yy}| = 1.725 \times 10^{-5}$ [m/Pa$\cdot$s]), we examine seven distinct strip locations with a starting location of $x_s=19.4$ [cm] up to $x_s=38.8$ [cm]. As the strip location is gradually moved downstream, we observe that each pair of phase values, $\theta_{Y_{yy}}$, which share intersecting $\Delta \mathcal{W}$ curves in Fig.~\ref{fig:vsweep}a exhibit a similar response to the change of strip location. Specifically, for each pair, the downstream PKEs relative to the reference case, $\Delta \tilde{\psi}$, appear to be far apart at lower $x_s$ values, slowly approach one another as $x_s$ increases, and eventually cross over. Hence, we choose in the following discussion to focus on a single pair, namely the two quadrature cases, $\theta_{Y_{yy}}=\pi/2$ and $\theta_{Y_{yy}}=3\pi/2$, which not only exhibit a crossing over of $\Delta \tilde{\psi}$, but also exhibit a major qualitative behavioral switch from downstream stabilization to destabilization, or vice versa. The other intersecting cases may therefore be interpreted as offset variations of this behavior.

As $x_s$ increases and the strip location shifts towards the downstream region, the work-rate term undergoes a significant amplitude change, as shown in the leftmost plot of Fig.~\ref{fig:vlocation}a. This figure depicts the maximum work-rate (signed peak value), $\Delta \mathcal{W}_{\mathrm{max}}$ within the strip region, for each strip location. The insets show the spatial work-rate profile, i.e., $\Delta \mathcal{W}(x)$, corresponding to the second strip location, $x_s=22.6$ [cm]. This notable amplitude change arises from the downstream pressure increase associated with the growth rate of the TS waves. Consequently, as the work-rate increases, the production term within the strip region is also affected, as shown in the middle plot of Fig.~\ref{fig:vlocation}a. This figure depicts the maximum production (signed peak value), $\Delta \mathcal{P}_{\mathrm{max}}$, within the strip region, for each strip location. The variations in the production are not limited to the strip region, but also extend into the contiguous region. The two insets in the figure show the spatial production profile, $\Delta \mathcal{P}(x)$ for the second ($x_s=22.6$ [cm]) and sixth ($x_s=35.6$ [cm]) strip locations. These spatial profiles exhibit an oscillatory behavior beyond the contiguous region, with the mean value of the oscillations changing as the strip location shifts downstream. The oscillatory behavior likely arises from the dipole-like $\Delta \mathcal{P}$ field seen in Fig.~\ref{fig:vlocation}b for $\theta_{Y_{yy}}=\pi/2$ (i.e., the two high-intensity contour regions). These changes in the mean value of the oscillatory production along with changes in the dissipation behavior, as a result of shifting the strip location, are the primary drivers of variations in the PKE, as will be detailed next.

The impact of the strip location on the downstream PKE reveals an intriguing behavior. The rightmost plot of Fig.~\ref{fig:vlocation}a depicts the maximum downstream PKE (signed peak value), $\Delta \tilde{\psi}_{\mathrm{max}}$, for each strip location. The maximum attenuation and amplification of the PKE caused by the strip for the phase values of $\theta_{Y_{yy}}=\pi/2$ and $\theta_{Y_{yy}}=3\pi/2$, respectively, take place at the most upstream strip location, i.e., $x_s=19.4$ [cm]. As $x_s$ increases, the PKE increases monotonically for $\theta_{Y_{yy}}=\pi/2$ and decreases monotonically for $\theta_{Y_{yy}}=3\pi/2$ up until a critical location at which the effect of the strip relative to the reference case becomes null ($x_s \approx 34$ [cm]). Further downstream of this location, the effect of the strip on the TS wave perturbation growth is reversed, with $\theta_{Y_{yy}}=\pi/2$ now destabilizing the wave and $\theta_{Y_{yy}}=3\pi/2$ stabilizing it. The two insets show the spatial PKE profile, i.e., $\Delta \tilde{\psi}(x)$, within the downstream region, corresponding to the second ($x_s=22.6$ [cm]) and sixth ($x_s=35.6$ [cm]) strip locations, which are approximately mirror images of one another.

The aforementioned changes in the TS wave behavior as $x_s$ increases largely arise from the combined effects of variations in production and dissipation and the decrease in the spatial window over which the PKE is allowed to grow or decay as a result of the flow-strip interactions, relative to the reference case. As can be seen in Fig.~\ref{fig:vlocation}b, the admittance strip generates a dipolar modulation pattern in the production field, when $\theta_{Y_{yy}}=\pi/2$. The dipole is oriented such that the pole associated with a production increase ($\Delta \mathcal{P} >0$, positive pole) is located slightly downstream of the one associated with a production decrease ($\Delta \mathcal{P} <0$, negative pole), with the negative pole being positioned closer to the wall. We also note that the exact opposite takes place when $\theta_{Y_{yy}}=3\pi/2$. While only shown for two $x_s$ values, this pattern similarly exists for all streamwise locations of the strip. Since the contour levels are normalized with respect to the maximum and minimum values (yielding a $[-1,1]$ scale), a comparison of the two cases shown in Fig.~\ref{fig:vlocation}b reveals a faster increase in the negative pole relative to the positive one, as indicated by the more faded appearance of the latter at the higher $x_s$ value. This is likely attributed to the negative pole being located nearer to the wall, rendering it more susceptible to the increased prescribed velocity imposed by the admittance strip, as a result of the elevated pressure, at the further downstream strip location. That said, the positive pole comprises a larger ``tail'' which extends further upward along the streamwise direction and spreads out more at farther downstream locations. This allows it to balance the increase in the negative pole on aggregate, or even become more dominant as the TS wave propagates from the contiguous to the downstream region, particularly because it is located close to the Reynolds-shear-stress peak. A close-up of the tail of the positive pole within the $\Delta \mathcal{P}$ plots is provided in Fig.~\ref{fig:vlocation}d for all seven strip locations.

The extended reach of the positive pole can be explained via the main components of the production term, namely the Reynolds shear stress, ${\tilde{u}\tilde{v}}$, and the wall-normal gradient of the streamwise base-velocity component, referred to as the shear rate, $\gamma=U_y$. Although the laminar boundary layer retains self-similarity, its thickness increases downstream, which in turn modifies the vertical profile of the shear rate, as illustrated in Fig.~\ref{fig:vlocation}c. As a result, the gradient becomes more vertically distributed downstream versus upstream where its more localized near the wall. In contrast, the Reynolds shear stress exhibits a pronounced peak at an approximately constant distance away from the wall, above which the correlation between the velocity components weakens. The rate of this de-correlation increases rapidly in the downstream direction, which escalates the drop in the Reynolds stresses, as shown in Fig.~\ref{fig:vlocation}c. As a result of this, shifting the strip downstream effectively allows the production components to support the positive pole, particularly its extended tail, more strongly than the negative pole. The latter is located closer to the wall where the base shear rate decreases significantly, whereas the former is positioned further from the wall where the shear rate reduction is weaker and may even increase at higher wall-normal levels. Furthermore, relative to their peak values, the Reynolds stresses undergo a substantial downstream reduction above the peak location in the wall-normal direction, which makes the disturbance induced by the strip in these upper layers more pronounced than its upstream counterpart. This can be interpreted as the new disturbance, caused by the downstream strip placement, exploiting the base energy available from the high shear rate, which cannot be fully digested by the weakened perturbation stresses. The amalgamation of these effects can be observed in the extended window past the strip, shown in Fig.~\ref{fig:vlocation}d, where the quantities are normalized by the peak value in each subplot. The increasing darker red-color spread, which is indicative of the growing positive pole tail, clearly corresponds to increasing $x_s$ values, i.e., shifting the strip location further downstream.  

A second key contributor to the PKE variations discussed earlier is dissipation, whose layer thickness is generally several times smaller than the boundary-layer thickness. As such, the dissipation layer is confined to the wall (near the negative production pole) and its thickness remains nearly uniform along the streamwise direction, in the order of $\sqrt{\mu_f/\pi\rho_f f}$ \cite{Michelis20232}. The variation in dissipation is shown in the bottom left panel of Fig.~\ref{fig:vlocation}b for a single case of $\theta_{Y_{yy}}=\pi/2$ and $x_s=22.6$ [cm], although the same pattern appears at the rest of the strip locations. The dissipation term, $-\Delta \mathcal{D}$, thus acts to weaken the negative production pole. The high-intensity dissipation spots extend downstream too, enabling the dissipation term to further limit the effect of the negative production pole. The net outcome of these counteracting terms is governed by the balance between $\Delta \mathcal{P}$ and $-\Delta \mathcal{D}$, which determines the resultant response. Specifically, the ratio of the vertically integrated positive-to-negative combination of production and dissipation, $|\Delta(\mathcal{P}-\mathcal{D})_{+}|/|\Delta(\mathcal{P}-\mathcal{D})_{-}|$, evaluated across the $x$-range covered by the contour plot, increases with strip location, $x_s$, as shown bottom right panel of Fig.~\ref{fig:vlocation}b, explaining the reversal in performance as the strip shifts downstream.

\begin{figure}[htbp]
\centerline{\includegraphics[width=0.5\columnwidth]{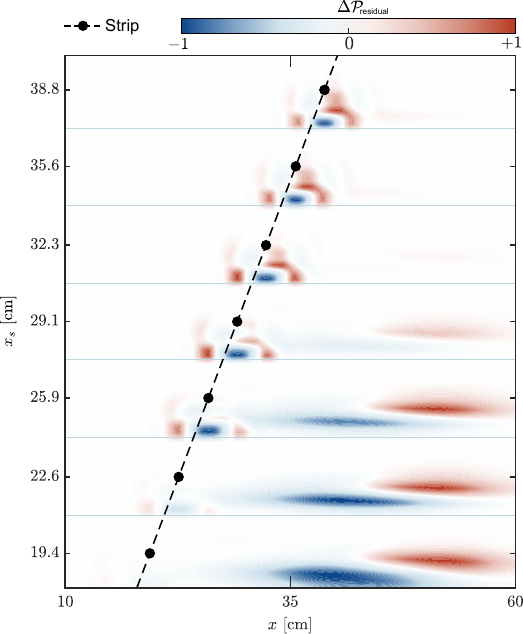}}
\caption{The non-oscillatory portion of the production variation, $\Delta\mathcal{P}_{\mathrm{residual}}$, obtained via Empirical Mode Decomposition (EMD) for the $\theta_{Y_{yy}}=\pi/2$ case. As $x_s$ increases, the positive production pole strengthens at the expense of the negative one, progressively diminishing the downstream dominant nature of the negative production until the two balance off. Following which, the contribution from the positive production pole becomes dominant downstream.}
    \label{EMD}
\end{figure}

\newpage
To further examine the impact of the strip location on the production, a further test is conducted, invoking the empirical mode decomposition (EMD) method \cite{Hao2022}. Due to the oscillatory nature of $\Delta \mathcal{P}$, the EMD method is well suited to elucidate the effect of $x_s$ by eliminating the first few intrinsic modes of the field and isolating the non-oscillatory portion of the production variation, $\Delta\mathcal{P}_{\mathrm{residual}}$. Figure~\ref{EMD} shows $\Delta\mathcal{P}_{\mathrm{residual}}$, which represents the background envelope, for all the considered strip locations when $\theta_{Y_{yy}}=\pi/2$. These envelopes tell the story: (1) At upstream strip locations (lower $x_s$ values), the negative production pole is able overcome the strength of its positive counterpart, leaving a trace which further amplifies downstream and effectively acts as destructive production that attenuates the downstream energy. (2) As the strip shifts downstream, the positive pole strength increases and becomes too much to surmount, hence reducing downstream stabilization. At a certain location, the effects of the two poles balance, reaching the null point around $x_s \approx 34$ [cm], where the strip causes no downstream effect relative to a rigid wall (observed earlier as the $\Delta \tilde{\psi}_{\mathrm{max}}$ crossing point in Fig.~\ref{fig:vlocation}a). Moving the strip further beyond this point gives the positive pole a dominant downstream advantage, leading to net destabilization.

Finally, we reiterate that the entire analysis in this subsection has been dedicated to the two quadrature cases, $\theta_{Y_{yy}}=\pi/2$ and $\theta_{Y_{yy}}=3\pi/2$, which is the only pair to exhibit a switch from downstream stabilization to destabilization. For the two remaining $\theta_{Y_{yy}}$ pairs that share intersecting $\Delta \mathcal{W}$ curves, i.e., $(3\pi/4,5\pi/4)$ and $(\pi/4,7\pi/4)$, the $\Delta \tilde{\psi}_{\mathrm{max}}$ curves still approach one another as $x_s$ increases and eventually intersect at what we refer to as the neutral point. Unlike the quadrature cases, however, this intersection is not associated with a stabilization-to-destabilization switch. In other words, the curves for each pair remain on one side of the zero axis despite intersecting at a shared $\Delta \tilde{\psi}_{\mathrm{max}}$ value. When this intersection coincides with $\Delta \tilde{\psi}_{\mathrm{max}}=0$, the neutral point becomes a null point, at which the strip imposes zero downstream effect on the flow, and the intersection therefore signals a major reversal from TS wave attenuation to attenuation, or vice versa, which is unique to the quadrature pair. Stated differently, all null points are neutral but not all neutral points are null. We also note that the location of the neutral point varies with the change in boundary conditions, as will be discussed later.  The in-phase ($\theta_{Y_{yy}} =0$) and out-of-phase ($\theta_{Y_{yy}} =\pi$) pair remain the only cases that do not share intersecting $\Delta \mathcal{W}$ curves, and therefore remain the extreme cases.

\subsubsection{Prescribed streamwise velocity location} \label{S_L}

\begin{figure}[htbp]
\centerline{\includegraphics[width=\columnwidth]{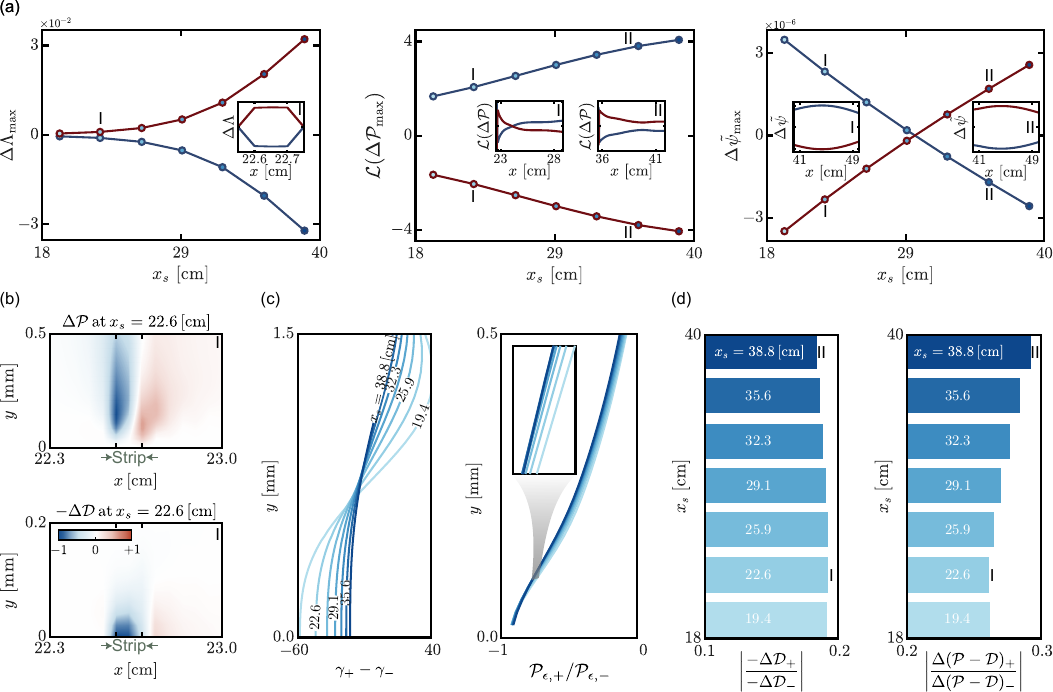}}
\caption{\textbf{ of strip location on TS wave growth in response to a prescribed streamwise velocity.} (a) Effect of changing the strip location, $x_s$, on the maximum signed values of the interfacial viscous energy, $\Delta {\Lambda}_{\max}$ (left), production, $\Delta {\mathcal{P}}_{\max}$ (middle), and the downstream PKE, $\Delta {\tilde \psi}_{\mathrm{max}}$ (right). The small insets show the spatial profiles of each of the three aforementioned parameters, for select $x_s$ values, from which the maximum values are extracted. (b) Production and dissipation levels compared to the reference case, $\Delta \mathcal{P}$ (top) and $-\Delta \mathcal{D}$ (bottom), corresponding to $x_s=22.6$ [cm]. (c) The variation along the wall-normal direction, $y$, in the shear-rate profile difference between the right and left edges of the strip, $\gamma_+ - \gamma_-$ (left), and the effective right-to-left production variation ratio, $\mathcal{P}_{\epsilon,+}/\mathcal{P}_{\epsilon,-}$ (right). (d) The effect of strip location on the positive-to-negative vertically integrated dissipation variations, $-\Delta \mathcal{D}$ (left), as well as the combined production and dissipation variations, $\Delta (\mathcal{P}-\mathcal{D})$ (right).}
\label{fig:ulocation}
\end{figure}

Turning to the influence of the location of a streamwise admittance strip on the TS wave growth, we utilize the same strip length ($\ell_s=1$ [mm]) and streamwise admittance ($|Y_{xx}| = 6.556 \times 10^{-4}$ [m/Pa$\cdot$s]), and relocate the strip between the same seven locations of Sec.~\ref{sec:W_L}. Unlike the wall-normal velocity scenario, the $\theta_{Y_{xx}}$ pairs which share intersecting $\Delta \Lambda$ curves do not encounter any crossing of downstream PKE levels as $x_s$ increases. However, the pairs which mirror each other around the zero line in Fig.~\ref{fig:usweep}a, e.g., $\theta_{Y_{xx}}=\pi/4$ and $\theta_{Y_{xx}}=3\pi/4$, do exhibit a neutral point at a different location from the one identified in the previous subsection. We will again focus here on the two cases which induce a switch between PKE stabilization and destabilization as the strip shifts downstream. For a streamwise admittance, these cases are $\theta_{Y_{xx}}=0$ and $\theta_{Y_{xx}}=\pi$. Similar to the previous analysis, the interface and production terms retain the same trend, both growing with $x_s$, as a result of the TS wave spatial amplification, as shown by $\Delta \Lambda_{\mathrm{max}}$ and $\Delta \mathcal{P}_{\mathrm{max}}$ in the leftmost and middle plots of Fig.~\ref{fig:ulocation}a, respectively. However, contrary to the wall-normal admittance scenario, the peak $\Delta \mathcal{P}_{\mathrm{max}}$ values within the strip and contiguous regions are not followed by oscillatory behavior, as confirmed by the spatial $\Delta \mathcal{P}$ profiles in the figure insets. The non-oscillatory behavior is a byproduct of the two production poles being situated at nearly the same vertical level, $y$, as shown in Fig.~\ref{fig:ulocation}b. The same figure reveals that negative pole on the left side spreads slightly further in the wall-normal direction, whereas the positive pole on the right extends further in the streamwise direction.

In terms of the impact of strip location on the downstream PKE, the $\Delta \tilde{\psi}_{\mathrm{max}}$ behavior depicted in the rightmost plot of Fig.~\ref{fig:ulocation}a preserves the overall trend attained by a wall-normal admittance, in that the maximum stabilization or destabilization is achieved when the strip is placed further upstream. The effect gradually decays as the strip is shifted downstream until reaching the null point, following which, the trend is reversed. While the process governing this performance flip is different from that of the wall-normal admittance, since, in the present case, the poles are nearly equidistant from the wall, it is still defined by the same two mechanisms, production and dissipation, along with their components, the base shear rate and the Reynolds shear stress. The base shear rate, $\gamma$, as shown previously in Fig.~\ref{fig:vlocation}c, experiences a gradient redistribution over the streamwise direction which results in a variation in the shear-rate difference around the strip. Figure~\ref{fig:ulocation}c presents the difference in the base shear rate between the right, $\gamma_+$, and left, $\gamma_-$, edges of the strip. Near the wall, the left pole exhibits a higher shear rate, with the difference between the two rates decreasing as the strip is moved downstream, particularly below the intersection point of the location shear rate curves, shown in Fig.~\ref{fig:ulocation}c, where the poles mainly resides. This reduction in the shear-rate difference indicates a potential for right pole dominance with the strip shifting downstream.

Since the base shear rate is just one component of production, the aforementioned potential for right pole dominance can be realized or missed depending on the second component, the Reynolds shear stress. The variation in the latter can be attributed, upon linearized approximation and ignoring the shear rate, to three terms: $|\tilde{u}_{\mathrm{Ref}}| |\delta_v| \cos \theta_{uv}$, $|\tilde{v}_{\mathrm{Ref}}| |\delta_u| \cos \theta_{uv}$, and $(|\tilde{u}_{\mathrm{Ref}}||\tilde{v}_{\mathrm{Ref}}| \sin \theta_{{u}{v}}) \Delta \theta$, with $\delta_{v}$ denoting the spatial disturbance of the wall-normal velocity induced by the prescribed streamwise velocity due to the inherent coupling of the velocity components (similar to the definition of $\delta_{u}$ in Sec.~\ref{Sec:S_P}). The first and second terms denote a perturbation in the wall-normal and streamwise velocities, respectively. The last term is the perturbation in the relative phase, $\theta_{uv}$, between the velocity components, referred to here as $\mathcal{P}_{\epsilon}$. The relative strengths of its positive and negative contributions are quantified using the ratio $\mathcal{P}_{\epsilon,+}/\mathcal{P}_{\epsilon,-}$, shown in Fig.~\ref{fig:ulocation}c, where $\mathcal{P}_{\epsilon,+}$ and $\mathcal{P}_{\epsilon,-}$ denote the positive and negative extrema, respectively, of the vertically integrated $\mathcal{P}_{\epsilon}$ distribution. This term is found to dominate and inherit the features of the original Reynolds shear stress, particularly above the dissipation layer. In particular, it shows superiority of the left pole over the right one as the strip shifts downstream. This also enhances the potential for right pole dominance provided by the base shear rate. The dominant pole changes as the the strip shifts downstream resulting in a performance swap between stabilization and destabilization. On the dissipation front, shown in the bottom of Fig.~\ref{fig:ulocation}b, two distinct high-intensity regions are observed as well. As the strip moves downstream, the positive region grows slower than the negative one, as indicated by their diminishing ratio as $x_s$ increases (Fig.~\ref{fig:ulocation}d). Here, similarly, $\Delta\mathcal{D}_{+}$ and $\Delta\mathcal{D}_{-}$ refer to the positive and negative extrema, respectively, of the vertically integrated dissipation perturbation field shown in Fig.~\ref{fig:ulocation}b. While this is in contrast with the behavior of the production poles, production remains dominant when both mechanisms are considered together. This is reflected by the increasing $|\Delta(\mathcal{P}-\mathcal{D})_+/\Delta(\mathcal{P}-\mathcal{D})_-|$ ratio as $x_s$ increases, as shown in Fig.~\ref{fig:ulocation}d. This explains the switch between TS-wave amplification and attenuation in the downstream PKE and confirms that production remains the primary mechanism governing this switch.

%%%%%%%%%%%%%%%%%%%%%%%%%%%%%%%%%%%%%%%%%%%%%%%%%%%%%%%%%%%%%%
\subsection{Admittance design rules and applicability to passive subsurface structures} \label{sec:SDR}

The influence of the admittance strip on TS wave growth was thoroughly investigated in Secs.~\ref{sec:A_P_Sweep} and \ref{sec:EB}, exploring the design space of the strip, i.e., phase, amplitude, location, and type of prescribed velocity at the fluid-structural boundary, along with the primary energy routes. Here, we summarize the key findings from the preceding sections into clear, impactful takeaways that turn detailed observations into actionable insights to guide the strip design. 

\begin{table*}[htpb]
\centering
\caption{Critical rules linking the strip interface response to downstream TS-wave behavior.}
\label{Tab:rules}
\small

\begin{tblr}{
    width=\textwidth,
    hspan=minimal,
    colspec={Q[c,m,wd=0.05\textwidth] X[0.7,l,m] X[1.2,l,m]},
    row{1}={font=\bfseries},
    rows={rowsep=4pt},
    hline{1,Z}={1pt},
    hline{2-Y}={0.45pt},
}

Rule
&
Strip interface response\textsuperscript{1}
&
Downstream implication
\\

\RuleBadge{ruleone}{1}
&
Entirely
\textcolor{destabcolor}{positive}
(\textcolor{stabcolor}{negative})\textsuperscript{2}
&
\textcolor{stabcolor}{Stabilization}
(\textcolor{destabcolor}{destabilization})
\\

\RuleBadge{ruletwo}{2}
&Crossing the zero axis\textsuperscript{3}
&
\textcolor{stabcolor}{Stabilization}-to-\textcolor{destabcolor}{destabilization} switch with shift in strip-location
\\

\RuleBadge{rulethree}{3}
&
Intersecting cases\textsuperscript{4}
&
PKE levels match at neutral strip location, reversing their relative ordering\textsuperscript{5}

\\

\RuleBadge{rulefour}{4}
&
Larger left-side response
&
\textcolor{wn}{Higher} (\textcolor{sw}{lower}) TS-wave control for the \textcolor{wn}{wall-normal} (\textcolor{sw}{streamwise}) boundary condition before the neutral point; reversed after

\\
\SetCell[c=3]{l,m}
\textbf{Takeaway:} Downstream PKE variation with strip location qualitatively resembles the interface response profile across the strip,
read left to right for \textcolor{wn}{wall-normal} and
right to left for \textcolor{sw}{streamwise} boundary conditions.
& &
\\
\end{tblr}

\vspace{2mm}

\begin{minipage}{0.98\textwidth}
\par
\raggedright
\footnotesize

\textsuperscript{1}The interface quantities used for the \textcolor{wn}{wall-normal} and \textcolor{sw}{streamwise} boundary conditions are \textcolor{wn}{$\Delta\mathcal{W}$} and \textcolor{sw}{$\mathcal{O}(\Delta\Lambda)$}, respectively.

\textsuperscript{2}No zero-line intersection.

\textsuperscript{3}The effective zero-line window is based on the strip width, slightly extended on both sides for the \textcolor{sw}{streamwise} boundary condition and slightly shifted to the left
for the \textcolor{wn}{wall-normal} one.

\textsuperscript{4}Intersections within the zero-line span.

\textsuperscript{5}Despite the reversal, the downstream variation attained at upstream strip locations is not recovered beyond the neutral point.
\end{minipage}
\end{table*}

\newpage
Despite the complex variations in PKE from the strip to the downstream region, the resulting downstream behavior, whether stabilization or destabilization, can be directly inferred from the interface terms, $\Delta \mathcal{W}$ or $\Delta \Lambda$, at the strip location. Table~\ref{Tab:rules} establishes four distinct design rules that translate the interfacial strip response into practical guidelines for selecting the phase and strip location required to achieve downstream attenuation or amplification of the TS wave energy.

In the following discussion, the interfacial strip response to a wall-normal admittance is represented by $\Delta\mathcal{W}$. For a streamwise admittance, we utilize the orthogonal component instead, $\mathcal{O}(\Delta \Lambda)=\frac{\mu_f}{\rho_f}\Im(\left.\tilde{u}_s\tilde{\xi} ^{*}\right|_{y=0})$, where $(.)^*$ denotes the complex conjugate, inducing the $\pi/2$ shift that is needed to generate the same downstream behavior as its wall-normal counterpart. With this treatment, the downstream PKE variations in response to the two boundary conditions could be tracked via the same set of rules. To formulate the criteria, we define an effective zero line as a virtual line spanning the strip width (e.g., $x_s=19.4$ to $19.5$ [cm] for the first strip location studied earlier), slightly extending past the strip boundaries to provide a margin for cases in which the interfacial response approaches the zero line near the strip edges. The first rule covers cases in which the interface response, or an extension of its trajectory, does not intersect this effective zero line. In such cases, an entirely positive response yields downstream stabilization, whereas an entirely negative response yields downstream destabilization. The second rule governs scenarios in which the interfacial response intersects the effective zero line. For both boundary conditions, such an intersection indicates that the downstream response switches between stabilization and destabilization after the null point as the strip location shifts downstream. Effectively, guaranteed downstream performance can be achieved, regardless of the strip location, when the strip operates away from the quadrature conditions (i.e., $\theta_{Y_{yy}}=\pi/2$ and $\theta_{Y_{yy}}=3\pi/2$) for a wall-normal admittance, and from the in-phase and out-of-phase conditions (i.e., $\theta_{Y_{xx}}=0$ and $\theta_{Y_{xx}}=\pi$) for a streamwise admittance, with the local response having zero possibility of intersecting the effective zero line.

The third rule applies to phase values which come in pairs, when the two interfacial responses of a pair intersect within the effective zero-line segment. When this happens, the downstream PKE of the pair relative to the reference case becomes equal at a neutral strip location. Following which, the two cases within the pair exchange performance, with the one case originally doing better upstream of the neutral point exhibiting a worse performance, and vice versa. However, we note that the highest attainable PKE variations with respect to the reference case, whether attenuation or amplification, exclusively take place when the strip is placed before the neutral point (similar to what was observed earlier in the rightmost panels of Figs.~\ref{fig:vlocation}a and \ref{fig:ulocation}a). Finally, the fourth rule pertains to the same pairs governed by the third rule, and serves to identify which phase value within each pair is favored in terms of its TS wave control ability, depending on the relative strip location with respect to the neutral point. The rule compares the values of the interface quantities at the starting (left) edge of the strip, $x_s$, henceforth termed the left-side response, for both cases within a pair, to determine which strip triggers a larger downstream PKE variation with respect to the reference case. For all strip locations preceding the neutral point, the phase value with the larger absolute left-side response induces a bigger downstream PKE change for a wall-normal boundary condition, and a smaller downstream PKE change for a streamwise boundary condition. The opposite takes place for strip located past the neutral point. The combination of these rules yields an intriguing observation: The variation of downstream PKE with strip location qualitatively reflects the spatial profile of the interface quantities across the strip, progressing from left to right for the wall-normal admittance and from right to left for the streamwise one.

\begin{figure}[htbp]
\centerline{\includegraphics[width=\columnwidth]{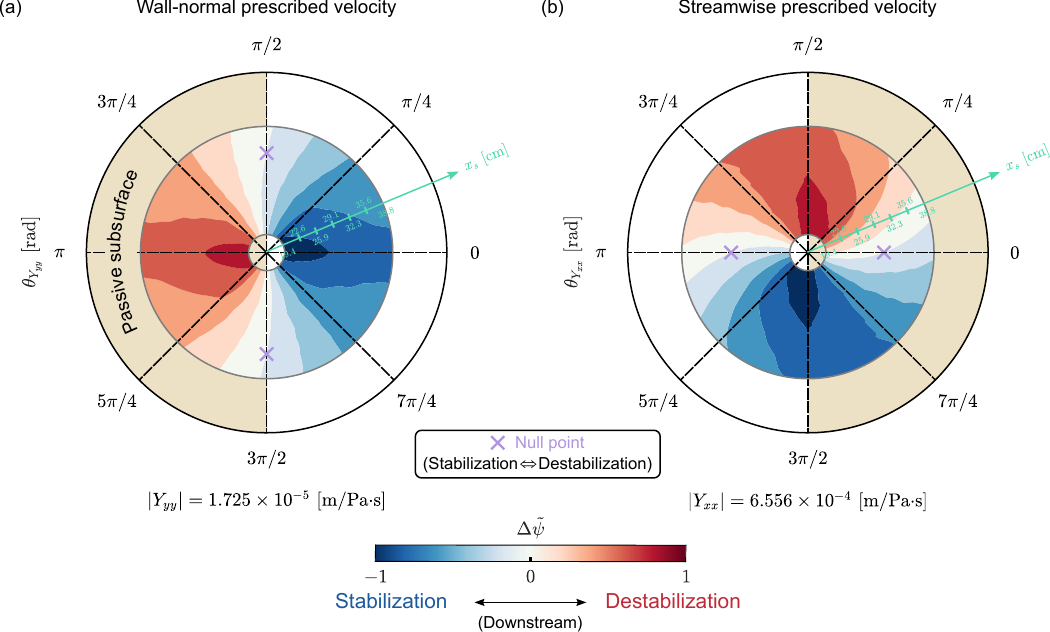}}
\caption{\textbf{Graphical representation of downstream stabilization rules.} (a) The left polar plot shows the variation of the downstream PKE, $\Delta \tilde{\psi}$ (contours), with the change in strip location, $x_s$ (radial axis), and the wall-normal admittance phase, $\theta_{Y_{yy}}$ (angular axis), in response to a wall-normal prescribed velocity. (b) The right polar plot shows the same variation versus the streamwise admittance phase, $\theta_{Y_{xx}}$, in response to a streamwise prescribed velocity. Corresponding admittance amplitudes are provided below each plot. Null points are indicated by crosses. They correspond to $\Delta \tilde{\psi}=0$, and mark the transition from downstream stabilization to destabilization, or vice versa. The design space physically realizable via a single passive subsurface is shaded for reference.}
\label{fig:Sub Design}
\end{figure}

Most of these rules can be visually identified by combining the admittance phases, strip location, and downstream PKE levels in graphical representation. Figure~\ref{fig:Sub Design} shows the variation in the normalized downstream PKE value, $\Delta \tilde{\psi}$, for both wall-normal and streamwise boundary conditions. The radial coordinate represents the strip location, $x_s$. The admittance amplitudes, $|Y_{yy}|$ and $|Y_{xx}|$, are the same as the ones used throughout the study. Regions in the vicinity of $\theta_{Y_{yy}}=\pi$ and $\theta_{Y_{xx}}=\pi/2$, which yield an entirely negative $\Delta\mathcal{W}$ and $\mathcal{O}(\Delta\Lambda)$, respectively, exhibit clear destabilization of the TS wave, as indicated by the overwhelmingly red region, and established by Rule 1. On the other hand, stabilization is evident in the neighborhood of $\theta_{Y_{yy}}=0$ and $\theta_{Y_{xx}}=3\pi/2$, as indicated by the overwhelmingly blue region. The bending of the contour edges along the radial direction, reminiscent of a fan blade twist, together with the appearance of the same contour level at different phases, reveals the mechanism underpinning Rule 3: Two phase values with comparable performance can evolve in opposite directions as the strip location shifts downstream, with the two cases sharing the same $\Delta \tilde{\psi}$ value at the neutral point. For a subset of cases, the same mechanism pushes the response through a null point, where the strip's effect on the downstream PKE becomes practically nonexistent (i.e., $\Delta \tilde{\psi}=0$, $\tilde{\psi}=\tilde{\psi}_{\mathrm{Ref}}$), marking a transition between stabilization and destabilization behavior with a shift in strip location (Rule 2). These points are marked with the crosses in Fig.~\ref{fig:Sub Design}. It is also worth noting that the bending of the contour edges is relatively stronger in response to a streamwise prescribed velocity (Fig.~\ref{fig:Sub Design}b), signaling that the neutral (and null) location is reached further upstream under such boundary condition that with a wall-normal one. This is consistent with the in-depth investigations shown in Figs.~\ref{fig:vlocation}a and \ref{fig:ulocation}a. 

Finally, we conclude this section with an important distinction between an elastic wall admittance and a full-fledged passive subsurface structure. As established earlier in Sec.~\ref{sec:FSI}, and again in Sec.~\ref{sec:A_P_Sweep}, the phase window realizable via a single passive subsurface is shaded in tan in Figs.~\ref{fig:Sub Design}a and b. These figures together with Figs.~\ref{fig:PKE admittance sweep}b and c highlight two paths to achieve concurrent local and downstream stabilization via a passive subsurface. The first is using a wall-normal phase between $\theta_{Y_{yy}}=\pi/2$ and $7\pi/12$. This narrow range shifts slightly with the streamwise location of the strip, as reflected by the fan-blade twist observed in Fig.~\ref{fig:Sub Design}. For example, for $\theta_{Y_{yy}}=\pi/2$, $x_s$ needs to be upstream of the null point to avoid the performance flip occurring at $x_s \approx 34$ [cm] (Rule 2). Alternatively, placing the strip further downstream after the null point while utilizing $\theta_{Y_{yy}}=3\pi/2$ can yield the desired downstream stabilization, albeit with a weaker outcome (refer to the footnote of Rule 3). However, in the context of an structure structure, and not an admittance strip, none of these paths could be achieved via a wall-normal deforming subsurface due to the emergence of a streamwise velocity component $\tilde{u}_s\neq0$, provided that proper coupling conditions are applied, as will be detailed later. The second path is through the design of a passive subsurface with a streamwise phase between $\theta_{Y_{xx}}=3\pi/2$ and $\theta_{Y_{xx}}=7\pi/4$. However, if robustness to changes in strip location, $x_s$, is an important design criteria, then $\theta_{Y_{xx}}\approx3\pi/2$ provides a safe, location-agnostic choice. In the next section, this route is successfully realized via a two-dimensional subsurface structure, demonstrating a pathway to achieving sustained downstream stabilization.

%%%%%%%%%%%%%%%%%%%%%%%%%%%%%%%%%%%%%%%%%%%%%%%%%%%%%%%%%%%%%%
\section{Fully coupled interactions and implications for phononic subsurface design} \label{sec:Fully coupled FSI}

The role played by an elastic strip embedded within the wall of a flat plate on the growth of TS waves, both in the vicinity of the strip and downstream of it, was studied in Sec.~\ref{sec:Parametric study}. The analysis in the previous section was conducted in the frequency domain using a prescribed velocity boundary condition, which enabled us to dissect the effect of the amplitude and phase of the strip's admittance function, as defined by Eq.~(\ref{eq:admittance BC}), and conduct a comprehensive energy budget analysis. In this section, we transition to fully-coupled simulations depicting fluid-structure interactions between a flow instability and a physically-defined subsurface structure, rather than an interface admittance. Studies are conducted using $\textsc{Comsol Multiphysics}$, implementing the Fluid Flow and Heat Transfer, the Acoustics, and the Structural Mechanics modules. To verify predictions from the previous section, we start by integrating a subsurface comprised of a one-dimensional phononic crystal, i.e., a phononic subsurface or PSub, which interacts with the flow through wall-normal vibrations. The latter is designed such that its frequency-dependent admittance amplitude falls within the moderate range identified in Sec.~\ref{sec:Parametric study}, roughly between $|Y_{yy}|=1\times10^{-4}$ and $|Y_{yy}|=1\times10^{-2}$ [m/Pa$\cdot$s], where the interfacing surface is neither an effective rigid wall or has an indefinite destabilizing effect on the TS waves. In Sec.~\ref{sec:1D PSub}, the performance of the PSub is directly compared to that of an equivalent admittance strip, and the results are complemented by the LST framework developed earlier. Following which, in Sec.~\ref{sec:physical realization of streamwise adm}, a two-dimensional PSub, or 2D-PSub, is conceptualized to invoke streamwise traction and incorporate the desired streamwise response, satisfying the second condition for downstream stabilization established in Sec.~\ref{sec:SDR}.

\subsection{One-dimensional PSub}
\label{sec:1D PSub}

\subsubsection{PSub design} \label{sec: 1D PSub Design}

The admittance strip from Sec.~\ref{sec:Parametric study} is replaced with a one-dimensional bi-layered PSub whose width along the streamwise direction spans the same $\ell_s=1$ [mm] distance, ensuring minimal spatial variation of the flow perturbation field across the structure (see Fig.~\ref{fig:PSub result}a). The PSub's unit cell is comprised of alternating layers, $A$ and $B$, chosen here to be PVC and Polyurethane foam, respectively, whose material and geometric properties are given in Table~\ref{Tab:1D PnC Mat prop}, with $\rho_i$, $E_i$, and $h_i$ denoting the density, elastic modulus, and height, respectively, and $i=A,B$ is an index denoting the corresponding layer. Consistent with the two-dimensional formulation, the cross-sectional area, $A_i$, is evaluated assuming a unit out-of-plane depth.

The PSub is an elastic bar which undergoes axial (i.e., longitudinal) vibrations only, allowing it to displace along the $y$-direction in response to the wall-normal traction imposed by the fluid, $\tilde{t}_{f_{yy}}=-\tilde{p}$. Prior to implementing a coupled fluid-structural study, the vibrational characteristics of the PSub are first obtained. Owing to its periodicity, the PSub exhibits Bragg scattering band gaps resulting in broad frequency windows in which impinging excitations are confined to the input location (in this case, the fluid-structural interface). As a result, the PSub's displacement, $\eta(y,t)$, within these band gaps are localized at the surface ($y=0$), enhancing its ability to influence the flow field directly above it. A transfer matrix, $\mathbf{T}$, can be defined which maps the displacement and forcing from one end of a unit cell to the next, such that $\mathbf{T}=\mathbf{T}_B \mathbf{T}_A$, where $\mathbf{T}_A$ and $\mathbf{T}_B$ are the transfer matrices of the individual layers, given by
\begin{equation}
\mathbf{T}_i
=
\begin{bmatrix}
\cos{(\kappa_i h_i)} & \frac{1}{z_i\omega}\sin{(\kappa_i h_i)}\\
{-z_i\omega}\sin{(\kappa_i h_i)} & \cos{(\kappa_i h_i)}
\end{bmatrix}
\end{equation}
where $\kappa_i={\omega}/{c_i}$ denotes the wavenumber within each layer, $c_i=\sqrt{E_i/\rho_i}$ is the speed of sound in each material, and $z_i=A_i\sqrt{E_i\rho_i}$ is the characteristic impedance of each layer \cite{al2017investigation}. The eigenvalues of $\mathbf{T}$ are considered a Floquet multiplier which can be used to extract the PSub's elastic wave propagation characteristics, by equating them to $e^{\mathrm{i}\tilde{\kappa}}$, where $\kappa$ is the elastic wavenumber, i.e., the spatial frequency of the structural wave, and $\tilde{\kappa}=\kappa~(h_A+h_B)$ is a non-dimensional form of it. The real part of this wavenumber, $\Re{(\tilde{\kappa})}$, determined as a function of frequency, produces the PSub's dispersion relation, which shows the structure's permissible frequency ranges for wave propagation (pass bands) and identifies band gaps as the intervals in between. Using the given parameters, the PSub is designed such that the TS wave spectrum falls within the first band gap. Figure~\ref{fig:PSub result}b depicts the PSub's dispersion diagram, showing a pass band up to $f=286$ [Hz], at which the first band gap starts. The TS wave range is marked by the dashed blue lines for convenience. 

\setlength\tabcolsep{1em}
\linespread{1.0}
\begin{table}[htb!]
\centering
\caption{Geometric and material properties of the one-dimensional bi-layered PSub.}
  \begin{tabular}{cccccc}
    \toprule
    \multirow{2}{*}{} &  Layer $A$ & Layer $B$  & units \\
      \midrule
        $\rho_i$ & $600$  & $60$ & $[\frac{\text{kg}}{\mathrm{m^3}}$]\\
		$E_i$ & $782$ &  $0.05$ & $[\text{MPa}$]\\
		$h_i$ & $10$  & $10$ & [$\text{mm}$]\\
    \bottomrule

  \end{tabular}
\label{Tab:1D PnC Mat prop}
\end{table}

The PSub used in this study is comprised of $10$ unit cells. It is anchored at the bottom and interfaces with the fluid at the top, as shown in Fig.~\ref{fig:PSub result}a. A purely structural analysis is conducted to quantify the PSub's resonant behavior, where a unit harmonic load, $|F| = 1$ [N/m$^2$], is imposed at the top, and the PSub's admittance function is obtained from $Y_{\mathrm{PSub},{yy}} = \mathrm{i}\omega \eta(0,\omega)/F$ via a frequency-domain analysis. In the FEM solver, Poisson's ratios for both materials are set to zero and streamwise displacement is suppressed to ensure a strictly one-dimensional response. By ensuring the TS wave spectrum falls within the first band gap, the PSub exhibits a reasonably wide non-resonant range in which a desired phase variation can be achieved via a single resonance intentionally placed inside the gap. Since the PSub consists of self-repeating unit cells, the starting and ending point of each cell do not influence the profile of an infinite chain but play an important role in the final (truncated) form of a finite one. Consequently, this in-gap resonance, commonly referred to as a truncation resonance, $f_{\mathrm{TR}}$, can be selectively placed at a given frequency by altering the unit cell configuration through a symmetry parameter, $\beta$ \cite{AlBabaa2023, AlBabaa2024}. In here, a PSub truncation mode at $f_{\mathrm{TR}}=495$ [Hz] is obtained by using $\beta = 0.875$. The amplitude and phase of $Y_{\mathrm{PSub},{yy}}$ are shown in Fig.~\ref{fig:PSub result}c, with the vertical and horizontal dashed lines indicating $f_{\mathrm{TR}}$ and the STP air admittance, respectively. For numerical stability, Rayleigh proportional damping is added to the structure, with a zero mass coefficient and a stiffness coefficient equal to  $\zeta/(\pi f_{\mathrm{TR}})$, where $\zeta = 1\times 10^{-4}$ is the modal damping ratio. Finally, we note that the PSub phase falls within the interval defined by $\theta_{Y_{\mathrm{PSub},{yy}}} \in [-\pi/2, \pi/2]$, consistent with Sec.~\ref{sec:A_P_Sweep}. As shown in the figure, the phase value switches between the extrema of this interval after resonance or anti-resonance peak, including at $f_{\mathrm{TR}}$.

\subsubsection{Limitations of flow control via one-dimensional PSub} \label{sec:1D F-PSubI}

\begin{figure*}[h!]
\includegraphics[width=1\textwidth]{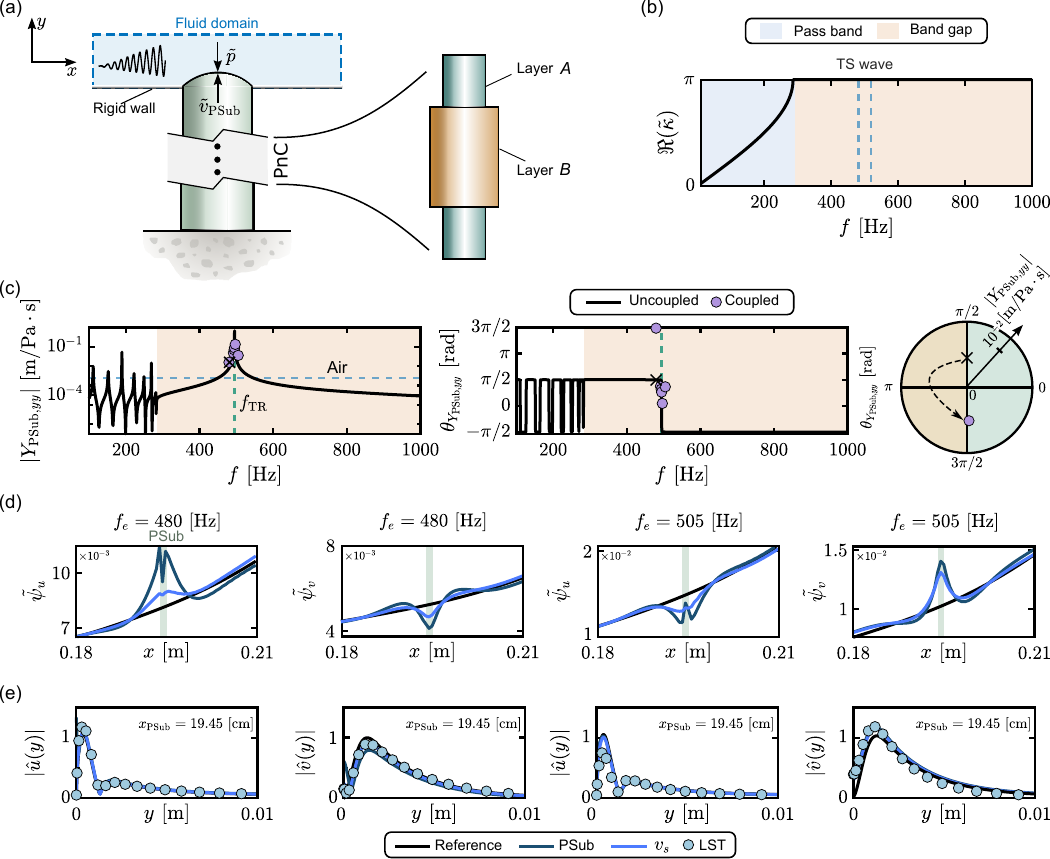}
    \caption{\textbf{Flow control via a single, one-dimensional passive PSub.} (a) An excerpt of the fluid domain with a TS wave instability interacting with a one-dimensional PSub, comprised of a bi-layered phononic crystal. (b) The PSub's dispersion diagram, with pass band and band gap regions shaded. Dashed vertical lines mark the frequency spectrum of the instability. (c) Frequency response functions of the PSub's wall-normal admittance amplitude and phase, $|Y_{\mathrm{PSub},yy}|$ and $\theta_{Y_{\mathrm{PSub},{yy}}}$. Black curves and circular markers denote uncoupled (purely structural) and coupled (fluid-structural) simulations, respectively. Horizontal and vertical dashed lines indicate air's nominal admittance at STP conditions and the truncation resonance frequency, $f_{\mathrm{TR}}$, respectively. The polar plot on the right highlights the phase shift between the uncoupled (cross) and coupled (circle) admittances for a representative case ($f_e=480$ [Hz]). (d) Streamwise and wall-normal PKE components, $\tilde{\psi}_u(x)$ and $\tilde{\psi}_v(x)$, for the reference (rigid-wall) case, the PSub case, and the case with a prescribed wall-normal velocity, $\tilde{v}_s$, applied via an admittance strip, at two distinct frequencies: $f_e = 480$ [Hz] (left) and $f_e=505$ [Hz] (right). (e) Streamwise and wall-normal mode shapes, |$\hat{u}(y)$| and |$\hat{v}(y)|$, extracted at $x_{\mathrm{PSub}} = 19.45$ [cm], for the three cases at the two same aforementioned frequencies.}
    \label{fig:PSub result}
\end{figure*}

After establishing the vibrational characteristics of the PSub, a fully coupled fluid-structure simulation is used to assess its effect on the flow perturbations. Two-way coupling is performed following Eq.~(\ref{eq:traction cont}), where the fluid's wall-normal traction excites the top surface of the PSub. Following which, the PSub's vibrational response at the interface is imposed on the fluid by ensuring continuity of the velocity field, following Eq.~(\ref{eq:velocity cont}). As explained earlier, a notable difference between results portrayed in this section and the generalized admittance boundary condition carried out in Sec.~\ref{sec:Parametric study} for an elastic strip is that the strip assumes no wall-normal deformation, leading to an isolated response of each velocity component, i.e., $\tilde{v}_s \neq 0$ with $\tilde{u}_s = 0$. In practice, however, the PSub deforms into the fluid leading to a coupling between the velocity components, which induces the linearized boundary conditions given by
\begin{subequations}
\begin{equation}
\tilde{u}_{\mathrm{PSub}} = -\eta(0,t) U_y 
\label{eq:FSI coupling u}
\end{equation}
\begin{equation}
\tilde{v}_{\mathrm{PSub}} = \frac{\partial \eta(0,t)}{\partial t}
 \label{eq:FSI coupling v}
\end{equation}
\label{eq:FSI coupling}
\end{subequations}
where $\tilde{u}_{\mathrm{PSub}}$ and $\tilde{v}_{\mathrm{PSub}}$ are the PSub's streamwise and wall-normal velocities along the interface boundary. To ensure the linear approximation in Eq.~(\ref{eq:FSI coupling u}) is valid, the surface deformation, $\eta(0,t)$, is constrained to be multiple orders less than the boundary layer displacement thickness, $\delta_1$. Simulation data of the perturbation field is extracted at the midpoint of the PSub, i.e., $x_{\mathrm{PSub}} = 19.45$ [cm], and a Fast Fourier Transform is applied to obtain the frequency spectrum of the acquired signals. The PSub's frequency-dependent admittance function at the interface, $Y_{\mathrm{PSub},yy}$, is then obtained. The amplitude and phase of $Y_{\mathrm{PSub},yy}$, as obtained from these coupled simulations for different TS-wave frequencies, is overlaid using the purple circular markers on top of the uncoupled function, plotted using the solid curves on the same plot in Fig.~\ref{fig:PSub result}c. Coupled values are shown for the narrow region between $480$ [Hz] and $505$ [Hz]. While the amplitudes of the coupled and uncoupled admittances show good matching in the vicinity of $f_{\mathrm{TR}}$, the phase values show expected differences between the aforementioned cases. Specifically, we focus on the pre-resonance forcing frequencies of $f_e=480$ [Hz], denoted in Fig.~\ref{fig:PSub result}c by a black cross. At this frequency, the pre-resonance $\theta_{Y_{\mathrm{PSub},{yy}}} = \pi/2$ phase value for the uncoupled case shifts to $\theta_{Y_{\mathrm{PSub},{yy}}} = 3 \pi /2$ once fluid-structural coupling takes shape. This difference is due to the negative sign in the fluid traction acting on the structure, $\tilde{t}_{f_{yy}}=-\tilde{p}$, as explained in footnote~\ref{FN:phase_shift}. In the frequency domain, this flipped sign leads to a $\pi$-shift in the phase angle between the uncoupled and coupled admittance functions (See footnote~\ref{FN:phase_shift} in Sec.~\ref{sec:A_P_Sweep}). The polar representation of the admittance function, shown in the rightmost panel of the same figure, illustrates this shift. The plot also confirms the matching amplitudes, $|Y_{\mathrm{PSub}}|$, between the uncoupled and coupled cases, as inferred from the near-equal radii of the cross and circular markers. 

In order to assess the impact of the PSub on TS wave growth inside the boundary layer, we re-write the PKE expression, Eq.~(\ref{eq:PKE}), to account for time-harmonic variation, yielding,
\begin{equation}
\tilde{\psi} = \frac{1}{2T_n} \, \rho_f \int_{0}^{y_{sl}} \int_{0}^{T_n} (\tilde{v}^2 + \tilde{u}^2) \,dt \,dy 
\label{eq:PKET}
\end{equation}
with $t_f=0.5$ [s] marks the total simulation time and $T_n$ denotes the periodic time of the perturbation over a sufficient number of cycles. Based on this, the streamwise and wall-normal components of the PKE, i.e., $\tilde{\psi}_u$ and $\tilde{\psi}_v$, respectively, are plotted versus the streamwise location, $x$, in Fig.~\ref{fig:PSub result}d for three distinct cases: The reference (rigid-wall) case, the PSub case, and the case with a prescribed wall-normal velocity, $\tilde{v}_s$, applied via an admittance strip, as detailed in Sec.~\ref{sec:Parametric study} (and ensured to have the same admittance amplitude and phase as the PSub at the selected frequencies for an accurate comparison). The variations in $\tilde{\psi}_u$ and $\tilde{\psi}_v$ for all three cases are shown at two different forcing frequencies, a pre-resonance frequency at $f_e = 480$ [Hz] and a post resonance one at $f_e = 505$ [Hz]. Inspecting these four plots, we observe the following: 1) The PSub and the prescribed wall-normal velocity cases show very similar overall patterns throughout the considered cases, confirming the efficacy and versatility of the generalized admittance boundary condition in predicting the behavior of a subsurface structure that matches the admittance parameters, regardless of the internal architecture of such structure. 2) Both the streamwise and wall-normal PKE components, $\tilde{\psi}_u$ and $\tilde{\psi}_v$, show completely opposite trajectories at one frequency versus the other. This is anticipated due to the flip in the admittance phase around the truncation resonance. 3) While the streamwise PKE, $\tilde{\psi}_u$, exhibits higher levels compared to the reference case at $f_e = 480$ [Hz] (i.e., destabilization), and lower levels at $f_e = 505$ [Hz] (i.e., stabilization) at the localized interaction region, the wall-normal PKE, $\tilde{\psi}_v$ shows the exact opposite behavior. 4) The amplitude of $\tilde{\psi}_u$ is notably larger than $\tilde{\psi}_v$ for both frequencies, rendering its individual effect to dominate the total PKE and therefore dictate the overall stabilization or destabilization outcome of the PSub (or admittance strip). As a result, the PSub achieves localized stabilization of the TS wave at $f_e=505$ [Hz], agreeing with previous reports \cite{Hussein2015, Kianfar2023-2, Barnes2021, Michelis2023}. 

While the results from the PSub and the prescribed wall-normal velocity condition via an admittance strip line up with respect to their local effect on the TS wave, i.e., in and around the region in which they directly interact with the flow, their effects on the same flow perturbation downstream appears to be different. This discrepancy is critically important since it brings to the forefront a fundamental limitation of using a single (passive) PSub to achieve broad TS wave attenuation. A closer look at Fig.~\ref{fig:PSub result}c reveals that the PSub-induced PKE reduction at $f_e=505$ [Hz], although larger locally than that of the equivalent strip, exhibits a stronger rebounding effect leading the downstream PKE to be even larger than that of the reference case. While the PKE corresponding to the prescribed velocity condition also recovers, its terminal value downstream remains closely below the reference level, effectively achieving strong localized attenuation while still maintaining modest downstream attenuation of the TS wave. The PSub's inability to achieve this dual combination is the byproduct of the streamwise velocity condition which, although not imposed, organically arises as a result of the wall-normal deformation, as captured by Eq.~(\ref{eq:FSI coupling u}). This emergent streamwise velocity contributes to a streamwise admittance which has the ability to shape the end result, especially given the dominant effect of the streamwise PKE on the overall outcome, as explained earlier. 

We also point out that this coupling-induced streamwise velocity condition associated with a PSub is not always unfavorable. For example, by inspecting the downstream effects observed at the pre-resonance forcing frequency of $f_e=480$ [Hz], it can be seen that the PSub successfully stabilizes the flow downstream, when the prescribed wall-normal velocity condition (i.e., the equivalent strip) fails to do so. This is also attributed to the same phenomenon. The strip simply follows the predictions set out in Fig.~\ref{fig:PKE admittance sweep}, because the wall-normal velocity is imposed in pure form, thereby avoiding any any unintended side effects along the streamwise direction. As a result, the wall-normal admittance parameters corresponding to this scenario, $|Y_{yy}|=0.0078$ [m/Pa$\cdot$s] and $\theta_{Y_{yy}}=4.7$ [rad], barely place the strip in the downstream destabilization region of Fig.~\ref{fig:PKE admittance sweep}b, confirming the downstream strip behavior seen in Fig.~\ref{fig:PSub result}d, of being just below the reference line at $f_e=480$ [Hz]. On the other hand, the PSub's ability to achieve downstream stabilization in this case is heavily nuanced. First, as explained in the previous paragraph, utilizing the PSub to impose a wall-normal velocity at the fluid-structural interface inevitably affects motion in the streamwise direction. In this particular case, this secondary effect can be quantified by extracting the effective streamwise admittance directly from the streamwise velocity and the shear stress at the interface. The results show a streamwise admittance of $|Y_{\mathrm{PSub},xx}|=6.02$ [m/Pa$\cdot$s] and $\theta_{Y_{\mathrm{PSub},xx}}=3.3$ [rad], which correspond to downstream stabilization (although outside the bounds of Fig.~\ref{fig:PKE admittance sweep}c)\footnote{The parameters used in this section are chosen such that the PSub's effect on the TS wave is pronounced and easy to visualize, relative to the reference case. Nevertheless, another set of results are provided in Appendix~\ref{app: MA PSub}, which illustrate the competing downstream effects of a PSub's coupled wall-normal and streamwise velocities, with admittance values that lie within the bounds of Fig.~\ref{fig:PKE admittance sweep}. \label{FN:moderate_amp}}. In other words, the wall-normal admittance (primary design target) and the streamwise admittance (secondary coupling effect) associated with the PSub trigger contradictory individual effects on the downstream TS wave. However, once again, given the significantly stronger influence that the streamwise PKE has on the aggregate behavior, the PSub in this case succeeds to achieve downstream stabilization, validating the outcome observed in Fig.~\ref{fig:PSub result}d.

Finally, we complement the PKE predictions with the mode shapes of the perturbation velocity components at the same two frequencies, for all three cases, as shown in Fig.~\ref{fig:PSub result}e. All mode shapes are extracted at $x_{\mathrm{PSub}} = 19.45$ [cm], and each mode shape is normalized by the maximum amplitude of the reference mode. At $f_e=480$ [Hz], the two leftmost panels of the figure show the streamwise velocity mode shape, $\hat{u}(y)$, for both the PSub and the prescribed wall-normal velocity condition to be slightly higher than the reference case (commensurate with localized destabilization), and vice versa for the wall-normal velocity mode shape, $\hat{v}(y)$ (localized stabilization). The exact opposite trend is observed at $f_e=505$ [Hz] in the two rightmost panels, consistent with the PKE predictions. In here, the results are further supplemented with the mode shapes obtained from the LST analysis. We note that, as expected, the LST results are shown to closely match the prescribed velocity rather than the PSub case due to the similarity in modeling, which allows both approaches to impose an isolated wall-normal velocity without accounting for any emergent streamwise velocity conditions that may arise as a result of coupling conditions. 

%%%%%%%%%%%%%%%%%%%%%%%%%%%%%%%%%%%%%%%%%%%%%%%%%%%%%%%%%%%%%%
\subsection{Two-dimensional PSub: A pathway toward downstream stabilization} \label{sec:physical realization of streamwise adm}

In Sec.~\ref{sec:1D F-PSubI}, we established that a one-dimensional PSub which vibrates in the wall-normal direction, and whose wall-normal admittance phase is $\theta_{Y_{\mathrm{PSub},yy}}= 3\pi/2$, can stabilize a TS wave when designed to operate on the appropriate side of the truncation resonance. However, contrary to prescribing an isolated wall-normal velocity at a strip of the wall, the PSub is simply incapable of achieving the twin benefits of a localized and downstream reduction in TS wave PKE levels relative to the reference case. The reason, as identified earlier, is influence of the PSub's wall-normal deformation on the fluid's streamwise velocity, defined by Eq.~(\ref{eq:FSI coupling u}). While the PSub, in theory, displaces the fluid in the wall-normal direction only, this coupling produces a de facto streamwise boundary condition whose effect cannot be ignored. 

It is important to note that the phase difference between the induced streamwise velocity and the interface shear stress cannot be independently controlled. Instead, it is influenced by several factors, including the phase relation between the perturbed pressure and wall shear stress, the structure's wall-normal displacement, and its coupling with the base-flow shear rate. This limitation directs attention to an alternative strategy in which the streamwise response is targeted directly, and its phase is controlled to concurrently produce local and downstream stabilization. Such a response can be achieved passively with a phase difference of $3\pi/2$ between the streamwise velocity and the interface shear stress, as demonstrated in Sec.~\ref{Sec:S_P} and shown in Fig.~\ref{fig:PKE admittance sweep}. An additional advantage of this approach is that purely streamwise structural motion does not introduce a wall-normal interface velocity, as proven in Appendix~\ref{app:2D BCs}, allowing the stabilizing effect of the streamwise boundary condition to be exploited without the competing influence of wall-normal motion. While, in theory, this can be achieved by adopting a one-dimensional structure that only vibrates in the streamwise direction, in practice, it is difficult to conceive of a one-dimensional structure that is completely fixed in the wall-normal direction (i.e., not responsive to pressure) but significantly responds to shear. Alternatively, a two-dimensional structure with extremely varying impedances (i.e., resistance to vibrational motion) in the two relevant directions can provide such pathway. Specifically, the structure can be designed to exhibit a significantly lower wall-normal than streamwise admittance, triggering a minuscule response in the wall-normal direction and promoting a predominantly streamwise response.

\begin{figure}[htbp]
\centerline{\includegraphics[width=1\columnwidth]{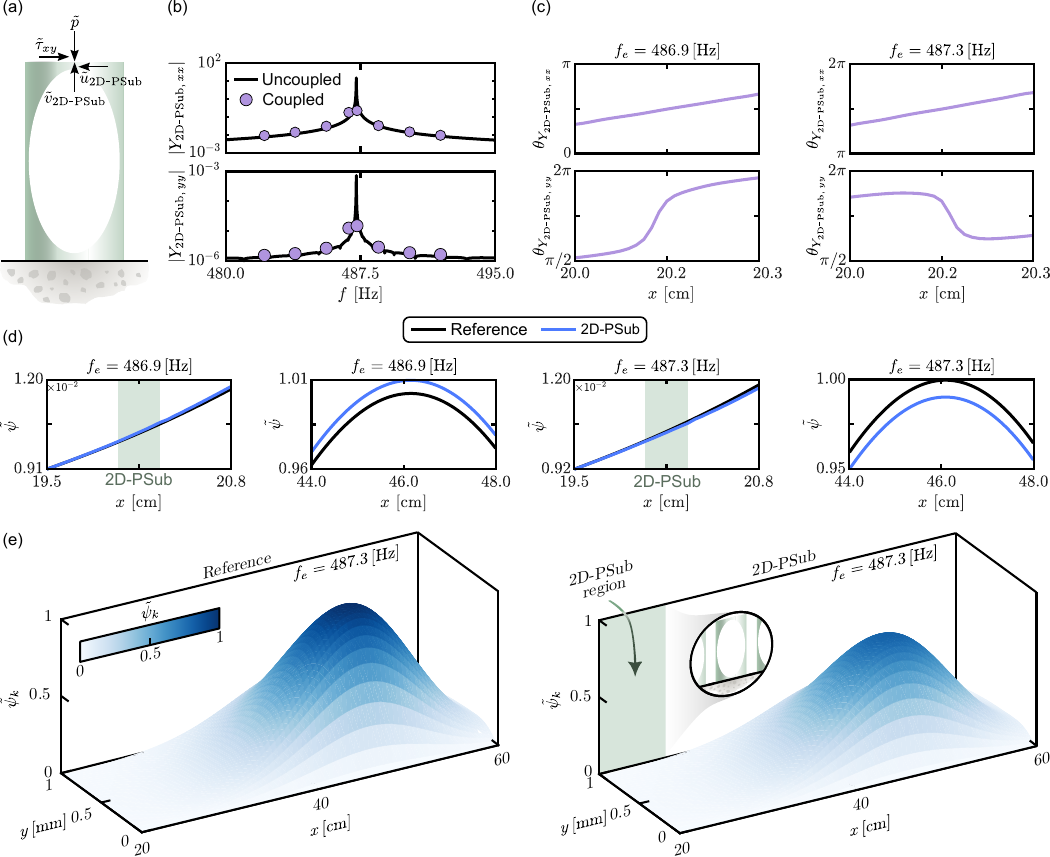}}
\caption{\textbf{Concurrent localized and downstream stabilization via an engineered streamwise admittance in a 2D-PSub.} (a) The 2D-PSub consists of a rectangular structure comprising an elliptical cavity. (b) Frequency response functions of the 2D-PSub's admittance in the streamwise and wall-normal directions, $|Y_{{\mathrm{2D\text{-}PSub}},xx}|$ and $|Y_{{\mathrm{2D\text{-}PSub}},yy}|$. Black curves and circular markers denote uncoupled (purely structural) and coupled (fluid-structural) simulations, respectively. (c) Spatial variation of the surface streamwise and wall-normal admittance phases, $\theta_{Y_{\mathrm{2D\mbox{-}PSub},xx}}$ and $\theta_{Y_{\mathrm{2D\mbox{-}PSub},yy}}$, over the length covered by the 2D-PSub at two frequencies, $f_e=486.9$ [Hz] (pre-resonance) and $f_e=487.3$ [Hz] (post resonance). (d) Variation of the total PKE along the streamwise direction, $\tilde\psi(x)$, at the two same frequencies for the 2D-PSub and reference (rigid-wall) cases. For each frequency, the left subplot shows the behavior at the 2D-PSub region and the right subplot shows the downstream trajectory. (e) The normalized pointwise PKE, $\tilde\psi_k(x,y)$, at $f_e=487.3$ [Hz] for the reference (left) and 2D-PSub (right) cases, adopting an array of $20$ 2D-PSubs, and showing approximately $20\%$ of PKE reduction in the latter relative to the former.}
    \label{fig:2D-PSub}
\end{figure}

We introduce a two-dimensional subsurface (2D-PSub) to validate the proposed strategy, providing a near-isolated and controllable streamwise velocity boundary condition. The 2D-PSub, shown in Fig.~\ref{fig:2D-PSub}a, is embedded within the rigid wall, occupying the streamwise interval $x_\mathrm{2D-PSub}\in[20,20.3]$~[cm], extends $6$ [mm] below the fluid surface, and is anchored at the bottom. The 2D-PSub is made of Polyethylene Terephthalate (PETE), whose density, $\rho_{\mathrm{2D\text{-}PSub}}$, elastic modulus, $E_{\mathrm{2D\text{-}PSub}}$, and Poisson's ratio, $\nu_{\mathrm{2D\text{-}PSub}}$, are listed in Table~\ref{Tab:rules}. An elliptical cavity is introduced within the subsurface to achieve the desired resonance frequency and tune the ratio between the streamwise and wall-normal admittance amplitudes. The minor and major axes of the elliptical cavity are chosen to be $24.3 \%$ and $22.5 \%$ of the 2D-PSub's length and height, respectively.

\setlength\tabcolsep{1em}
\linespread{1.0}
\begin{table}[htb!]
\centering
\caption{Material properties of the 2D-PSub.}
  \begin{tabular}{cccccc}
    \toprule
    \multirow{2}{*}{} &  PETE & units \\
      \midrule
        $\rho_{\mathrm{2D\text{-}PSub}}$ & $2200$ & $[\frac{\text{kg}}{\mathrm{m^3}}$]\\
		$E_{\mathrm{2D\text{-}PSub}}$ & $400$ & $[\text{MPa}$]\\
		$\nu_{\mathrm{2D\text{-}PSub}}$ & $0.4$  &  [-]\\
    \bottomrule

  \end{tabular}
\label{Tab:2Dprop}
\end{table}

In a purely structural analysis, the 2D-PSub is subjected to the same uncoupled frequency-domain studies described in Sec.~\ref{sec:1D F-PSubI}, with the goal of computing its uncoupled admittance frequency response functions at the top surface. The streamwise and wall-normal admittance amplitudes, $|Y_{{\mathrm{2D\text{-}PSub}},xx}|$ and $|Y_{{\mathrm{2D\text{-}PSub}},yy}|$, are shown in Fig.~\ref{fig:2D-PSub}b, for a narrow window of TS-wave frequencies, $f_e \in [480,495]$ [Hz], which encompasses a resonance at $487.2$ [Hz]. The amplitude ratio, $\left| Y_{{\mathrm{2D\text{-}PSub}},xx}/ Y_{{\mathrm{2D\text{-}PSub}},yy} \right|$, is shown to be sufficiently large ($\approx 10^4$) across the considered range, which compensates for the fact that the fluid wall-normal traction, $\tilde{p}$, is greater than the streamwise counterpart, $\tilde{\tau}_{xy}$, as shown in Fig.~\ref{fig:Comp Setup}d. As a result, the streamwise response constitutes the only effective boundary condition at the interface, relative to a wall-normal response which is practically nonexistent. The circular markers in Fig.~\ref{fig:2D-PSub}b represent the admittance amplitudes extracted from the coupled simulations. These are obtained by averaging the perturbation field and the 2D-PSub response over the length of the fluid-structural interface, and computing the coupled admittances accordingly. This explains the slight deviation from the uncoupled curves due to due to the spatial variation in the fluid's traction. The upper panel of Fig.~\ref{fig:2D-PSub}c shows the spatial variation of the streamwise admittance phase, $\theta_{Y_{\mathrm{2D\text{-}PSub},xx}}$, at two frequencies, before ($f_e=486.9$ [Hz]) and after ($f_e=487.3$ [Hz]) resonance. As expected, $\theta_{Y_{\mathrm{2D\text{-}PSub},xx}}$ crosses $\pi/2$ at the midpoint of the 2D-PSub at the pre-resonance frequency, and crosses $3\pi/2$ at the same location at post resonance. Recall that the latter scenario was shown to achieve the favorable combination of local and downstream stabilization in Sec.~\ref{Sec:S_P}. On the other hand, the lower panel of the same figure shows the spatial variation of the wall-normal admittance phase, $\theta_{Y_{\mathrm{2D\text{-}PSub},yy}}$, at the two same frequencies. In contrast to the two $\theta_{Y_{\mathrm{2D\text{-}PSub},xx}}$ plots which showed the same trajectory at both frequencies, $\theta_{Y_{\mathrm{2D\text{-}PSub},yy}}$ exhibits a noticeable variation in how it changes over the length of the 2D-PSub, depending on the frequency of interest.

After establishing the vibrational characteristics of the 2D-PSub, Eqs.~\eqref{eq:FSI coupling} are modified to account for both the streamwise and wall-normal motions of the subsurface. Unlike the one-dimensional PSub, the elastic deformation of the 2D-PSub varies spatially in both directions, with $\eta(x,y,t)$ and $\lambda(x,y,t)$ denoting the wall-normal and streamwise displacement components, respectively. Accounting for these variations at the fluid-structure interface yields (refer to Appendix~\ref{app:2D BCs} for details)
\begin{subequations}
\begin{equation}
\tilde{u}_{\mathrm{2D\text{-}PSub}} =\frac{\partial \lambda(x_\mathrm{2D-PSub},0,t)}{\partial t} - \eta(x_\mathrm{2D-PSub},0,t) U_y
\label{eq:2D FSI coupling u}
\end{equation}
\begin{equation}
\tilde{v}_{\mathrm{2D\text{-}PSub}} =\frac{\partial \eta(x_\mathrm{2D-PSub},0,t)}{\partial t}
\label{eq:2D FSI coupling v}
\end{equation}
\label{eq:2D FSI coupling}
\end{subequations}
Figure~\ref{fig:2D-PSub}d compares the PKE of the reference case to that of the 2D-PSub case both in the vicinity of the 2D-PSub region and downstream. At the pre-resonance frequency of $f_e=486.9$ [Hz], the 2D-PSub intervention with the flow is detrimental in both regions, leading to higher $\tilde{\psi}$ levels than the reference case both locally and downstream. However, post resonance at $f_e=487.3$ [Hz], the 2D-PSub generates lower $\tilde{\psi}$ levels than the reference case at the interaction region in addition to a pronounced and sustained attenuation at larger $x$ values, thus achieving the elusive combination of localized and downstream stabilization. For further verification, an array of 20 independent 2D-PSubs, each spanning the same $3$ [mm] length are lined up along the streamwise direction with a $1$ [mm] gap in between (resulting in a total coverage of $80$ [mm]), corresponding to a TS wave frequency of $f_e=487.3$ [Hz]. The pointwise PKE, $\tilde{\psi}_k$, is presented in Fig.~\ref{fig:2D-PSub}e for both the reference and 2D-PSub cases. For fair comparison, $\tilde{\psi}_k$ in both plots is normalized by the global maximum pointwise PKE of the reference case. Very encouragingly, the array of 2D-PSubs produces significant attenuation of the TS-wave within the fluid domain, resulting in nearly $20\%$ reduction in $\tilde{\psi}_k$ relative to the reference case.

Finally, we emphasize that although the 2D-PSub considered in this analysis is comprised of a single material, the underlying design principle is not limited to this specific configuration. The presented framework allows for any class of architected materials, including phononic crystals and locally resonant metamaterials, to be similarly employed provided that the structure interfacing with the flow exhibits a sufficiently large ratio of streamwise to wall-normal admittance. Maintaining this directional anisotropy is essential to preserve the desired attenuation response while preventing undesirable amplification, thereby paving the way for a broader design framework for realizing effective and streamwise-sustained passive flow control via engineered subsurface deformations.

%%%%%%%%%%%%%%%%%%%%%%%%%%%%%%%%%%%%%%%%%%%%%%%%%
\section{Concluding remarks} \label{sec:conclusion}

In this work, we revisited the problem of passive flow stabilization, building upon the foundational principle of perturbation energy attenuation through synchronized wall deformations. The LNS equations were utilized to simulate the growth of TS waves within the boundary layer, at a forcing frequency of $f_e = 500$ [Hz]. Both the high-fidelity FEM and the supporting LST analyses successfully demonstrated moderate spatial growth, showing excellent agreement, and confirming the reliability of the established framework for capturing linear TS wave growth. We investigated the influence of an elastic strip embedded within a localized section of a flat plate on the growth of a TS wave instability in a wall-bounded flow. The analysis examined the flow field both in the immediate vicinity of the strip and further downstream, where the PKE achieves its maximum growth.

The design space of the elastic strip was comprehensively explored using a generalized admittance matrix model, by imposing prescribed streamwise and wall-normal velocity boundary conditions over the strip region. A systematic parametric sweep encompassing phase, amplitude, and strip location for each velocity component independently enabled the isolation of each parameter's effect on the interference mechanism with the underlying instability. A moderate strip admittance amplitude was found to be most effective, whereas extremely low and high admittances had no effect or became unconditionally destabilizing across the entire phase spectrum, respectively. Examining various locations of flow-strip interaction, upstream strip placement proved most effective for establishing strong TS wave interactions, whereas moving the strip downstream was shown to induce an intriguing performance reversal at specific admittance phases. Nevertheless, the admittance phase still emerged as the critical tuning parameter for TS wave stabilization. Two distinct pathways were identified for simultaneous local and downstream attenuation using a passive admittance strip: A wall-normal admittance phase satisfying $\theta_{Y_{yy}} \in [\pi/2, 7\pi/12]$, or a streamwise admittance phase in the range $\theta_{Y_{xx}} \in [3\pi/2, 7\pi/4]$, with the understanding that the precise bounds of these phase ranges may shift slightly with the streamwise location of the strip. While the wall-normal condition was shown to be unrealizable via a single, passive wall-normal deforming subsurface due to implicit coupling between the different velocity components in true operating conditions, the streamwise condition offered a viable pathway for downstream stabilization using a passive, streamwise deforming subsurface. A complementary energy budget analysis elucidated the strip's impact on the different energy routes, revealing that the identified phase intervals favorably modified the dominant mechanisms governing downstream energy production and dissipation. 

Motivated by the objective of achieving simultaneous stabilization both locally and downstream, the streamwise phase interval identified from the admittance strip analysis was used to guide the design of an optimized structure. In order to achieve a tailored streamwise-dominant response, a two-dimensional phononic subsurface (2D-PSub) was conceptualized. The 2D-PSub was designed to exhibit a wall-normal admittance significantly lower than its streamwise counterpart, thereby suppressing the wall-normal response to a negligible level and promoting predominantly streamwise motion. In pursuit of this goal, an embedded elliptical cavity within a rectangular structure was utilized to fine tune the amplitude ratio between the streamwise and wall-normal deformations, while preserving control over the structure's resonant frequency of interest. Our results demonstrated that an array of these 2D-PSubs successfully achieves dual stabilization of TS waves, yielding a $20 \%$ reduction in the perturbation kinetic energy of the flow and establishing a promising pathway for passive boundary-layer flow control.

%%%%%%%%%%%%%%%%%%%%%%%%%%%%%%%%%%
\section*{Acknowledgments}
The authors acknowledge support of this work by the US Air Force Office of Scientific Research (AFOSR) under award no. FA9550-23-1-0564.

%%%%%%%%%%%%%%%%%%%%%%%%%%%%%%%%%%
\bibliographystyle{unsrt} % Or your preferred style
\bibliographystyle{apsrev4-2-titles_up_2}
\bibliography{references}

%%%%%%%%%%%%%%%%%%%%%%%%%%%%%%%%%%
\clearpage
\appendix
\renewcommand{\thesection}{\Alph{section}}
\renewcommand{\thesubsection}{\thesection-\arabic{subsection}}

\makeatletter
\renewcommand{\@seccntformat}[1]{\csname the#1\endcsname\quad}
\makeatother
\setcounter{figure}{0} 
\renewcommand{\thefigure}{A\arabic{figure}}
\setcounter{table}{0} 
\renewcommand{\thetable}{A.\arabic{table}} 

\section{Appendix}
\subsection{Finite element meshing protocol}
\label{app:Mesh}

Two meshes are considered for the current framework, with the first and second meshes respectively corresponding to the first and second steps in Sec.~\ref{sec:Num Frame}. For the first step which resolves the boundary layer profile over the flat plate, the smallest element height in the first mesh is $h_{\mathrm{min,1}}=1\times10^{-5}$ [m] which is two orders of magnitude less than the largest boundary layer displacement thickness $\delta_1(x=1) = 1.4\times10^{-3}$ [m]. The element height is increased along the wall-normal direction, starting at the wall ($y=0$ [m]), and ending at the beginning of the sponge layer ($y_{sl}=0.1$ [m]), reaching a maximum element height of $h_{\mathrm{max,1}} = 2\times10^{-3}$ [m] at the top of the domain. The sponge layers have an average element height of $h_{sl} = 5.5\times10^{-3}$ [m]. The element width inside the entire computational domain, including the sponge layer, is $\ell_{m,1} = 1$ [mm].

\setlength\tabcolsep{1em}
\linespread{1.0}
\begin{table}[htb!]
\centering
\caption{Number of mesh elements and corresponding velocity amplitude error at peak TS wave growth location}
  \begin{tabular}{ccccc}
    \toprule
    \multirow{2}{*}{} $r$ & $N_e$ & $e_{\tilde{u}}$ [$\%$] & $e_{\tilde{v}}$ [$\%$]  \\
      \midrule
      $1$ & $136166$  & $-$ & $-$\\
		 $2$ & $238960$ &  $59.93 $ & $59.08 $\\
		 $3$& $266800$ & $30.81 $ &  $31.21 $\\
		 $4$ & $444960$  & $0.47 $ & 
         $0.57 $\\
    \bottomrule
  \end{tabular}
\label{Tab: Mesh conv}
\end{table}

The second step solves for the perturbation field, permitting it to afford a larger element height compared to the first step while ensuring the TS wavelength near the wall is captured by an adequate number of elements. The minimum height is changed to $h_{\mathrm{min,2}}=2\times10^{-5}$ [m] while the maximum height becomes $h_{\mathrm{max,2}}=2\times10^{-2}$ [m] at the top of the sponge layer ($y=0.3$ [m]). The second mesh element width is set to $\ell_{m,2} = 0.5$ [mm]. This ensures that there is at least 20 elements per wavelength along the streamwise direction, given the TS wave frequencies considered in this study (The average TS wavelength is around $1.7$ [cm]).  A mapping step is performed in $\textsc{Comsol}$ ensuring that the boundary layer from the first mesh is conformed to the second mesh. Additionally, the structural mesh is conformed to the fluid mesh at the interaction boundary by setting the width of the structural elements to $\ell_{m,2}$ with a structural element height of $h_{\mathrm{struc}}=1.4\times10^{-3}$ [m]. 

A mesh convergence study is performed on the second mesh by varying the element height and width in the fluid domain. The total number of mesh elements, given by $N_e$, is then computed. We observe the error in magnitude of the velocity components at the peak TS wave growth location near the wall for each iteration, given by $e_{\tilde{\mathbf{u}}} =(\left|\tilde{\mathbf{u}}(r+1)|-|\tilde{\mathbf{u}}(r)|\right)/|\tilde{\mathbf{u}}(r+1)|$, where $r$ is the iteration number. Table~\ref{Tab: Mesh conv} lists the percentage errors for the streamwise and wall-normal velocities, $e_{\tilde{u}}$ and $e_{\tilde{v}}$, respectively, corresponding to each $N_e$.

%%%%%%%%%%%%%%%%%%%%%%%%%%%%%%%%%%%%%%%%%%%%%%%%%%
\subsection{Velocity boundary conditions}
\label{app:VBC}

In Sec.~\ref{sec:Parametric study}, the elastic segment of the wall interacting with the flow is modeled as an admittance strip via a generalized admittance boundary condition. To facilitate the parametric study, a prescribed velocity boundary condition was utilized by explicitly imposing the amplitude and phase of each velocity component by brute force. This method is assessed here against the prescribed admittance boundary condition set by Eq.~(\ref{eq:admittance BC}), where the amplitude and phase of the admittance terms, $Y_{yy}$ and $Y_{xx}$, are defined beforehand and multiplied by the averaged corresponding traction terms over the strip to obtain the strip velocity. As such, the prescribed velocity represents a mathematical form of the boundary condition that is agnostic to the strip properties, whereas the admittance boundary condition encodes the strip design by incorporating its elastic admittance, $\mathbf{Y}$, from which the velocity can be obtained through the appropriate traction. Regardless of the way this boundary condition is inputted, the admittance values outputted from the simulation, i.e., extracted from the fluid response through the perturbation field parameters, should reasonably match. 

\begin{figure*}[h!]
\includegraphics[width=1\textwidth]{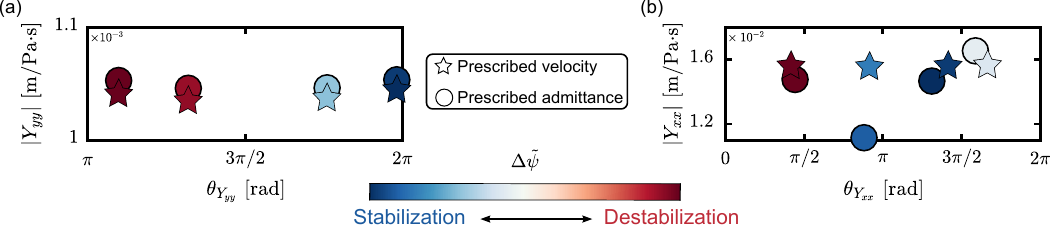}
    \caption{Variations in the perturbation kinetic energy relative to the reference case, $\Delta \tilde{\psi}$, at the strip location, with changes in (a) wall-normal admittance amplitude and phase, $|Y_{yy}|$ and $\theta_{Y_{yy}}$, and (b) streamwise admittance amplitude and phase, $|Y_{xx}|$ and $\theta_{Y_{xx}}$. Star and circular markers correspond to prescribed velocity and prescribed admittance boundary conditions, respectively.}
    \label{fig:Velocity BC verf}
\end{figure*}

The obtained admittances at the middle of the strip are presented in Fig.~\ref{fig:Velocity BC verf}. The figure shows the amplitude versus phase of the wall-normal and streamwise admittances,  $Y_{yy}$ and $Y_{xx}$, for the two aforementioned boundary condition approaches (star markers for prescribed velocity and circular markers for prescribed admittance). The PKE level at the strip, $\Delta \tilde{\psi}$ is denoted by the color of the marker itself, showing the degree of stabilization or destabilization, relative to the reference (rigid-wall) case. For both boundary conditions the admittance is extracted by averaging the corresponding traction and velocity across the strip region. For $Y_{yy}$, the two approaches show negligible differences in the obtained amplitudes and perfect alignment in terms of the PKE (Fig.~\ref{fig:Velocity BC verf}a). For $Y_{xx}$, a slight variation in the amplitude and phase appears (Fig.~\ref{fig:Velocity BC verf}b). However, the matching of the PKE levels across all considered cases confirms that the stabilization or destabilization pattern is still well captured by the prescribed velocity boundary condition.

%%%%%%%%%%%%%%%%%%%%%%%%%%%%%%%%%%%%%%%%%%%%%%%%%%
\subsection{Perturbation kinetic energy equation}
\label{app:PKE FD}

To derive the PKE equation in the frequency domain, we start with the velocity components at a given streamwise location, denoted by $\mathbf{\tilde{u}}= \Re( \mathbf{\hat{u}}(y) e^{\mathrm{i}\omega t})$. The PKE is obtained from the velocity components, i.e.,
\begin{equation}
   \tilde{\psi} =\frac{1}{2}\rho_{{f}}\int_0^{y_{sl}}\left[{\tilde{v}}^2 + \tilde{u}^2\right] dy 
\end{equation}
where $\tilde{v}^2=\frac{1}{4} \left(\hat{v}(y)e^{\mathrm{i}\omega t}+ \hat{v}(y)^*e^{\mathrm{-i}\omega t}\right)^2$ and likewise for $\tilde{u}^2$. Expanding the squared term and taking a time average over one cycle, the aforementioned identity can be simplified to $\tilde{v}^2=\frac{1}{2} \hat{v}(y) \hat{v}(y)^*$. Finally, by using $\hat{v}(y) \hat{v}(y)^* = |\hat{v}(y)|^2$ and omitting the wall-normal dependence $(y)$ for brevity, we arrive at the frequency domain PKE expression, given by
\begin{equation}
   \tilde{\psi} =\frac{1}{4}\rho_{{f}}\int_{0}^{y_{sl}}\left[|{\hat{v}}|^2 + {|\hat{u}|}^2\right] dy 
\end{equation}

%%%%%%%%%%%%%%%%%%%%%%%%%%%%%%%%%%%%%%%%%%%%%%%%%%
\subsection{Energy budget equation}
\label{app:Energy_budget}

By expanding the convective terms in Eq. \eqref{eq:LNS} and decomposing the equation into its two directional components, we arrive at
\begin{subequations}
\begin{equation}
\frac{\partial \tilde u}{\partial  t} + U\tilde u_x + V\tilde u_y + \tilde u\,U_x + \tilde v\,U_y
= -\frac{1}{\rho_f}\,(\tilde p_x) + \frac{\mu_f}{\rho_f} \,(\nabla^2 \tilde u)
\label{eq:ulin}
\end{equation}
\begin{equation}
\frac{\partial \tilde v}{\partial  t} + U\tilde v_x + V\tilde v_y + \tilde u\,V_x + \tilde v\,V_y
= -\frac{1}{\rho_f}\,(\tilde p_y) + \frac{\mu_f}{\rho_f} \,(\nabla^2 \tilde v)
\label{eq:vlin}
\end{equation}
\end{subequations}
Multiplying Eq.~\eqref{eq:ulin} by $\tilde u$ and Eq.~\eqref{eq:vlin} by $\tilde v$, and adding the two outcomes, yields
\begin{align}
\tilde u \frac{\partial \tilde u}{\partial  t} + \tilde v \frac{\partial \tilde v}{\partial  t}
+ U(\tilde u \tilde u_x + \tilde v \tilde v_x)
+ V(\tilde u \tilde u_y + \tilde v \tilde v_y)
+ \tilde u^2 U_x
+ \tilde u \tilde v\,U_y
+ \tilde u \tilde v\,V_x
+ \tilde v^2 V_y
\nonumber\\
=
-\frac{1}{\rho_f}\left(\tilde u\,\tilde p_x+\tilde v\,\tilde p_y\right)
+\frac{\mu_f}{\rho_f}\left(\tilde u\nabla^2\tilde u+\tilde v\nabla^2\tilde v\right)
\label{eq:pre_energy}
\end{align}
Using the pointwise specific PKE allows the relevant terms to be collected into the following compact form,
\begin{align}
\tilde u \frac{\partial \tilde u}{\partial  t} + \tilde v \frac{\partial \tilde v}{\partial  t} = \frac{\partial \tilde \psi_k}{\partial t},\ \ \ \
\tilde u \tilde u_x + \tilde v \tilde v_x = \frac{\partial \tilde \psi_k}{\partial x},\ \ \ \
\tilde u \tilde u_y + \tilde v \tilde v_y = \frac{\partial \tilde \psi_k}{\partial y}
\end{align}
and, therefore, Eq.~\eqref{eq:pre_energy} becomes
\begin{align}
\frac{\partial \tilde \psi_k}{\partial t}
+U\frac{\partial \tilde \psi_k}{\partial x}
+V\frac{\partial \tilde \psi_k}{\partial y}
+\tilde u^2 U_x
+\tilde u\tilde v\,U_y
+\tilde u\tilde v\,V_x
+\tilde v^2 V_y \nonumber\\
=-\frac{1}{\rho_f}\left(\tilde u\,\tilde p_x+\tilde v\,\tilde p_y\right)
+\frac{\mu_f}{\rho_f}\left(\tilde u\nabla^2\tilde u+\tilde v\nabla^2\tilde v\right)
\label{eq:mid_energy}
\end{align}

\noindent
Using the product rule and the perturbation continuity equation given by Eq.~\eqref{eq:Continuity2}, we get
\begin{align}
-\frac{1}{\rho_f}\left(\tilde u\,\tilde p_x+\tilde v\,\tilde p_y\right)
=
-\frac{1}{\rho_f}\bigg(\frac{\partial(\tilde u\tilde p)}{\partial x}+\frac{\partial(\tilde v\tilde p)}{\partial y}\bigg)
\label{eq:pressure_transport}
\end{align}

\noindent
Moreover, for two-dimensional incompressible flow, the perturbation vorticity can be used to express the viscous contribution as follows
\begin{align}
\left(\tilde u\nabla^2\tilde u+\tilde v\nabla^2\tilde v \right) =-\bigg(\frac{\partial(\tilde u\tilde \xi)}{\partial y}
-\frac{\partial(\tilde v\tilde \xi)}{\partial x}
+\tilde \xi^2 \bigg)
\label{eq:viscous_identity_dim}
\end{align}
Therefore, substituting Eqs.~\eqref{eq:pressure_transport} and \eqref{eq:viscous_identity_dim} into Eq.~\eqref{eq:mid_energy} yields
\begin{align}
\frac{\partial \tilde \psi_k}{\partial t}
+U\frac{\partial \tilde \psi_k}{\partial x}
+V\frac{\partial \tilde \psi_k}{\partial y}
+\tilde u^2 U_x
+\tilde u\tilde v\,U_y
+\tilde u\tilde v\,V_x
+\tilde v^2 V_y
+\frac{1}{\rho_f}\frac{\partial(\tilde u\tilde p)}{\partial x} \nonumber\\
+\frac{1}{\rho_f}\frac{\partial(\tilde v\tilde p)}{\partial y}
=
-\frac{\mu_f}{\rho_f}\bigg(
\frac{\partial(\tilde u\tilde \xi)}{\partial y}
-\frac{\partial(\tilde v\tilde \xi)}{\partial x}
+\tilde \xi^2
\bigg)
\label{eq:final_local_dim}
\end{align}
\noindent
Eq.~\eqref{eq:final_local_dim} is the pointwise PKE equation for a general two-dimensional incompressible base flow, which can also be rewritten, using the base continuity equation in Eq.~\eqref{eq:Continuity}, as follows
\begin{align}
\frac{\partial \tilde \psi_k}{\partial t}
+\frac{\partial}{\partial x}\Big(U\tilde \psi_k+\frac{\tilde u\tilde p}{\rho_f}-\frac{\mu_f}{\rho_f} \tilde v\tilde \xi\Big)
+\frac{\partial}{\partial y}\Big(V\tilde \psi_k+\frac{\tilde v\tilde p}{\rho_f}+\frac{\mu_f}{\rho_f} \tilde u\tilde \xi\Big)
=\nonumber\\
-\left(
\tilde u^2 U_x
+\tilde u\tilde v\,U_y
+\tilde u\tilde v\,V_x
+\tilde v^2 V_y
\right)
-\frac{\mu_f}{\rho_f} \tilde \xi^2
\label{eq:conservative_dim}
\end{align}
\noindent
For a time-periodic perturbation, the time average, denoted by $\overline{(\cdot)}$, of the time-derivative term in Eq.~\eqref{eq:conservative_dim} tends to zero. Integrating from the wall ($y=0$) to a far-field location ($y=y_\infty$) allows further reduction. Perturbation quantities are assumed to decay to zero at $y_\infty$, and the base flow satisfies $U=V=0$ at the wall, while the perturbation remains consistent with the wall or interface disturbance. These assumptions together yield the final one-dimensional budget equation,
\begin{equation}
\begin{aligned}
\frac{d}{dx}
\left[
\overbrace{\int_0^{y_\infty} U\overline{{\tilde{\psi}}_k}\,dy}^{(\text{I})}
+\overbrace{\int_0^{y_\infty}\frac{1}{\rho_{{f}}} \overline{\tilde u\tilde p}\,dy}^{(\text{II})}
-\overbrace{\int_0^{y_\infty}\frac{\mu_{{f}}}{\rho_{{f}}} \overline{\tilde v\tilde \xi}\,dy}^{(\text{III})}
\right]
=
\underbrace{-\int_0^{y_\infty}\overline{\tilde u\tilde v}\,U_y\,dy}_{(\mathcal{P})}
-
\\[4pt]
\underbrace{\int_0^{y_\infty} \frac{\mu_{{f}}}{\rho_{{f}}}\overline{\tilde \xi^2}\,dy}_{(\mathcal{D})}
-\underbrace{\int_0^{y_\infty}\overline{\tilde u\tilde v}\,V_x\,dy}_{(\text{i})}
+\underbrace{\int_0^{y_\infty}
\left(\overline{\tilde v^2}-\overline{\tilde u^2}\right)\,U_x\,dy}_{(\text{ii})}
+\underbrace{\left. \frac{1}{\rho_{f}}\overline{\tilde v_s\tilde p} \right|_{y=0}}_{(\mathcal{W})}
+\underbrace{\left. \frac{\mu_{{f}}}{\rho_{{f}}} \overline{\tilde u_s\tilde \xi}\right|_{y=0}}_{(\Lambda)}
\end{aligned}
\label{appeq:Energy Budget}
\end{equation}
In the frequency-domain formulation, the $\overline{(\cdot)}$ notation is omitted and the quadratic time-averaged quantities are evaluated from their complex values through
\begin{equation}
    \tilde{a}\tilde{b}=\frac{1}{2}\Re({\hat{a}\hat{b}^{*}}) \quad , \quad
    \tilde{a}^{\,2} = \frac{1}{2}\left|\hat{a}\right|^{2}
\end{equation}

%%%%%%%%%%%%%%%%%%%%%%%%%%%%%%%%%%%%%%%%%%%%%%%%%%
\subsection{Phase Computation}
\label{app:phase_comp}

For a complex perturbation $\tilde{q}=\Re({\tilde{q}})+\mathrm{i}\Im({\tilde{q}})$, the phase is defined as $\arg (\tilde{q}) = \operatorname{atan2}\big(\Im(\tilde{q}),\Re(\tilde{q})\big)$. In here, the representative phase is obtained by averaging the complex value across the fluid-structural interface, such that
\begin{equation}
\theta_q=
\arg\bigg(\frac{1}{N}\sum_{j=1}^{N}\tilde{q}(x_j,0)\bigg)
\end{equation}
\noindent
The velocity and pressure are evaluated from the first grid point while the streamwise velocity gradient is calculated using the following second-order one-sided difference
\begin{equation}
\tilde{u}_y(x,0)
\approx
\frac{-3\tilde{u}_1+4\tilde{u}_2-\tilde{u}_3}
{2(y_2-y_1)}
\end{equation}
where the subscripts $1$, $2$, and $3$ denote the first, second, and third grid points, respectively, in the wall-normal direction. Since $\tilde{\tau}_{xy}(x,0)=\mu_f\,\tilde{u}_y(x,0)$, the two quantities have identical phases. Accordingly, the phase of the wall-normal admittance, $\theta_{Y_{yy}}$, is computed from the phase difference between the wall-normal velocity and the pressure, whereas the phase of the streamwise admittance, $\theta_{Y_{xx}}$, is computed from the phase difference between the streamwise velocity and the shear stress.

%%%%%%%%%%%%%%%%%%%%%%%%%%%%%%%%%%%%%%%%%%%%%%%%%%
\subsection{Downstream TS wave stabilization effect of a single PSub} 
\label{app: MA PSub}

In Sec.~\ref{sec:1D F-PSubI}, the individual PKE components, $\tilde{\psi}_u$ and $\tilde{\psi}_v$, of the PSub were shown at two excitation frequencies, $f_e=480$ [Hz] and $f_e=505$ [Hz], resulting in four distinct plots. Focusing on the two $\tilde{\psi}_u$ plots, since the influence of this PKE component on the final control outcome was deemed to be dominant, it can be observed that the PSub achieves downstream TS wave stabilization at $f_e=480$ [Hz], relative to the reference case, and the exact opposite at $f_e=505$ [Hz]. In discussing the underlying mechanisms behind these contrasting downstream effects, it was pointed out that although the PSub introduces wall-normal motion into the fluid, the coupling between the velocity components results in streamwise motion, which effectively gives rise to a streamwise admittance. The amplitude of the latter was computed to be $|Y_{\mathrm{PSub},xx}| = 6.02$ [m/Pa$\cdot$s] and $|Y_{\mathrm{PSub},xx}| = 2.32$ [m/Pa$\cdot$s] at $f_e = 480$ [Hz] and $f_e=505$ [Hz], respectively. These values fall outside of the moderate amplitude region conveyed by Fig.~\ref{fig:PKE admittance sweep}c, but they were nevertheless chosen because they provide a tangible effect on the TS wave, compared to the reference case, allowing for optimal visualization of the PSub's effect. In here, for completeness, we provide the results from an additional PSub study, where the material parameters, $E_i$ and $\rho_i$, given in Sec.~\ref{sec:1D F-PSubI} are scaled up by a factor of $S=70$, to lower the induced streamwise admittance values to a range which can be readily tracked using Fig.~\ref{fig:PKE admittance sweep}c, allowing us to align the predicted outcomes from this figure with the PKE behavior achieved from the coupled PSub simulations. We note that scaling the properties ensures that the modified design has the same wave propagation characteristics, preserving the bandgap range and the resonant frequencies, and allowing us to examine the PSub under the two same frequencies.

\begin{figure*}[h!]
\includegraphics[width=1\textwidth]{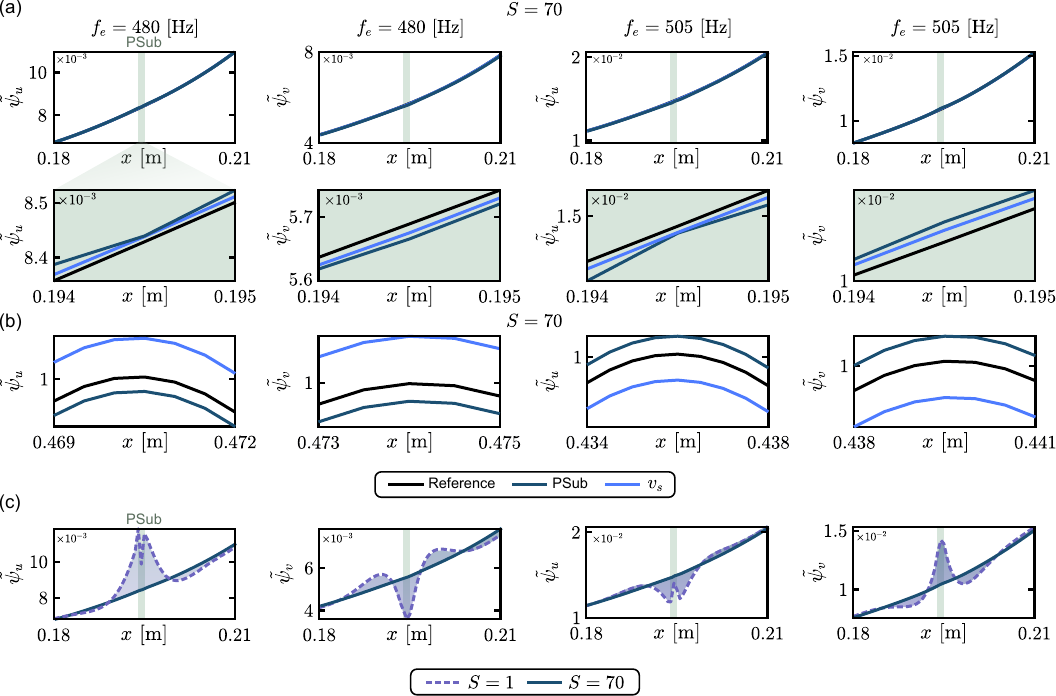}
    \caption{(a) A moderate amplitude PSub case, $S=70$, with the streamwise and wall-normal normalized PKE components, $\tilde{\psi}_u(x)$ and $\tilde{\psi}_v(x)$, for the reference (rigid-wall) case, the PSub case, and the case with a prescribed wall-normal velocity, $\tilde{v}_s$, applied via an admittance strip, at two distinct frequencies: $f_e = 480$ [Hz] (pre-resonance) and $f_e=505$ [Hz] (post resonance). Bottom panel shows the zoomed inset at the strip. (b) The same PKE components at the same frequencies at the downstream region. (c) The PKE components, for the two PSub cases, $S=1$ and $S=70$.}
    \label{fig:PSub additional study}
\end{figure*}

The extracted streamwise admittance amplitudes from the simulations of the scaled PSub are found to be $|Y_{\mathrm{PSub},xx}| = 8.4\times10^{-2}$ [m/Pa$\cdot$s] and $|Y_{\mathrm{PSub},xx}| = 1.1\times10^{-1}$ [m/Pa$\cdot$s], at $f_e = 480$ [Hz] and $f_e=505$ [Hz], respectively. Figures~\ref{fig:PSub additional study}a and b show the PKE components, $\tilde{\psi}_u$ and $\tilde{\psi}_v$, for the reference (rigid-wall) case, the PSub case, and the case with a prescribed wall-normal velocity, $\tilde{v}_s$, applied via an admittance strip, at the vicinity of the PSub and downstream of it. Due to the lowered admittance, the effect of the PSub on the PKE levels can best be observed in the close-up insets provided. At $f_e=480$ [Hz] (pre-resonance), and contrary to the admittance strip, the PSub achieves downstream stabilization, as shown in the leftmost plot of Fig.~\ref{fig:PSub additional study}b. The PSub's streamwise admittance phase for this case is $\theta_{Y_{\mathrm{PSub},xx}} = 5.5$ [rad], which is clearly in the downstream stabilization zone of Fig.~\ref{fig:PKE admittance sweep}c, confirming the PSub's behavior. On the other hand, at $f_e=505$ [Hz] (post resonance), and again contrary to the admittance strip, the PSub causes the TS waves to destabilize downstream. The PSub's streamwise admittance phase for this case is $\theta_{Y_{\mathrm{PSub},xx}} = 2.4$ [rad], which falls squarely within the downstream destabilization zone of Fig.~\ref{fig:PKE admittance sweep}c, confirming the ability of the predictive approach charted in Sec.~\ref{sec:A_P_Sweep} to forecast the PSub's expected influence on the TS wave downstream of the interaction location. 

Finally, Fig.~\ref{fig:Velocity BC verf}c, compares the performance of the scaled (i.e., $S=70$) and the original (i.e., $S=1$) PSub by showing their effect on the different PKE components on the same plot. Despite the similarity in the overall qualitative outcomes, the figure shows the stark difference between the degree to which each PSub alters the PKE level of the TS wave, relative to the reference case, as a result of the difference in admittance amplitudes.

%%%%%%%%%%%%%%%%%%%%%%%%%%%%%%%%%%%%%%%%%%%%%%%%%%
\subsection{Boundary conditions of the 2D-PSub at the fluid-strutural interface}
\label{app:2D BCs}

To evaluate the boundary conditions of the 2D-PSub associated with the interface motion, we examine the displacement of the interface as a result of a change in the 2D-PSub's location from an initial position $(x_{\mathrm{2D-PSub}},0)$ to a new position $(x_{\mathrm{2D-PSub}}+\lambda,\eta)$, at a given time, $t$. Since the total velocity is decomposed into a base velocity, $\mathbf{u}=(U,V)$, and a perturbation, $\tilde{\mathbf{u}}=(\tilde{u},\tilde{v})$, the displacement of the 2D-PSub leads to the following boundary condition
\begin{equation}
\mathbf{u}(x_\mathrm{2D-PSub}+\lambda,\eta)+\mathbf{\tilde{u}}(x_\mathrm{2D-PSub}+\lambda,\eta)=\bigg(\frac{\partial \lambda}{\partial t},\frac{\partial \eta}{\partial t}\bigg)
\label{eq:intereq}
\end{equation}
A Taylor expansion around the initial position results in
\begin{subequations}
\begin{equation}
    \mathbf{u}(x_\mathrm{2D-PSub}+\lambda,\eta)=\mathbf{u}(x_\mathrm{2D-PSub},0)+\lambda\mathbf{u}_x(x_\mathrm{2D-PSub},0)+\eta\mathbf{u}_y(x_\mathrm{2D-PSub},0)+.....
    \label{eq:Te1}
\end{equation}
\begin{equation}
    \mathbf{\tilde{u}}(x_\mathrm{2D-PSub}+\lambda,\eta)=\mathbf{\tilde{u}}(x_\mathrm{2D-PSub},0)+\lambda\mathbf{\tilde{u}}_x(x_\mathrm{2D-PSub},0)+\eta\mathbf{\tilde{u}}_y(x_\mathrm{2D-PSub},0)+.....
        \label{eq:Te2}
\end{equation}
\end{subequations}
where higher-order terms of $\lambda$ and $\eta$ are truncated. By substituting Eqs.~\eqref{eq:Te1} and \eqref{eq:Te2} into Eq.~\eqref{eq:intereq}, and separating streamwise and wall-normal components, we arrive at the following equations
\begin{subequations}
\begin{equation}
    U+\tilde{u}+\lambda\left(U_x+ \tilde{u}_x\right)+\eta\left(U_y+ \tilde{u}_y\right)=\frac{\partial \lambda}{\partial t}
    \label{eq:bcx1}
\end{equation}
\begin{equation}
V+\tilde{v}+\lambda\left(V_x+ \tilde{v}_x\right)+\eta\left(V_y+ \tilde{v}_y\right)=\frac{\partial \eta}{\partial t}
    \label{eq:bcy1}
\end{equation}
\end{subequations}
where all quantities evaluated at $(x_\mathrm{2D-PSub},0)$. Eqs.~\eqref{eq:bcx1} and \eqref{eq:bcy1} can be reduced further. Since the wall is rigid and stationary in the reference case, the base flow satisfies no-slip and no-penetration, i.e., $U(x_{\mathrm{2D-PSub}},0)=V(x_{\mathrm{2D-PSub}},0)=0$, which also leads to $U_x(x_{\mathrm{2D-PSub}},0)=V_x(x_{\mathrm{2D-PSub}},0)=0$. Additionally, applying the base-flow continuity equation gives $V_y(x_{\mathrm{2D-PSub}},0)=0$. Finally, we note that $\lambda,\eta,\tilde{u}$, and $\tilde{v}$ are all treated as small quantities of the same order, enabling second and higher-order terms to be neglected. This reduces Eqs.~\eqref{eq:bcx1} and \eqref{eq:bcy1} to
\begin{subequations}
\begin{equation}
    \tilde{u}+\eta U_y=\frac{\partial \lambda}{\partial t}
    \label{eq:bcx2}
\end{equation}
\begin{equation}
    \tilde{v}=\frac{\partial \eta}{\partial t}
    \label{eq:bcy2}
\end{equation}
\end{subequations}
and, as a result, the boundary conditions can be written as
\begin{equation}
\tilde{u}_{\mathrm{2D\text{-}PSub}} =\frac{\partial \lambda(x_\mathrm{2D-PSub},0,t)}{\partial t} - \eta(x_\mathrm{2D-PSub},0,t) U_y
\label{eqapp:2D FSI coupling u}
\end{equation}
\begin{equation}
\tilde{v}_{\mathrm{2D\text{-}PSub}} =\frac{\partial \eta(x_\mathrm{2D-PSub},0,t)}{\partial t}
\label{eqapp:2D FSI coupling v}
\end{equation}

\end{document}